\documentclass[a4paper,11pt]{article}
\usepackage{jheppub} 
\usepackage{lineno}

\usepackage{amssymb}
\usepackage{amsmath}
\usepackage{color}
\usepackage{xspace}
\usepackage{appendix}
\usepackage{float}
\usepackage{hyperref}
\usepackage{diagbox}
\usepackage{multirow}
\usepackage{multicol}
\usepackage{subfig}
\usepackage{comment}
\usepackage{soul}
\usepackage[compat=1.1.0]{tikz-feynhand}
\allowdisplaybreaks
\allowdisplaybreaks

\title{Next-to-next-to-leading order threshold soft function for $tW$ production}
\author[a]{Jia-Le Ding}
\author[a]{,~Hai Tao Li}
\author[a,b]{, and Jian Wang}

\affiliation[a]{School of Physics, Shandong University, Jinan, Shandong 250100, China}
\affiliation[b]{Center for High Energy Physics, Peking University, Beijing 100871, China}

\emailAdd{dingjl@mail.sdu.edu.cn}
\emailAdd{haitao.li@sdu.edu.cn}
\emailAdd{j.wang@sdu.edu.cn}

\abstract{
We compute the two-loop soft function for the associated production of a top quark and a  $W$ boson near the threshold, where the invariant mass of the $tW$ system approaches the collider energy. 
We employ the reverse unitarity technique and integration-by-parts identities to reduce soft loop integrals to a minimal basis of master integrals. 
These master integrals are analytically evaluated using the method of differential equations, yielding results expressed in terms of multiple polylogarithms. Additionally, we analyze the asymptotic behavior of the soft function in the low and high energy limits.  Our results provide a vital component for threshold resummation at the next-to-next-to-next-to-leading logarithmic level in this process.
}

\begin{document}
\maketitle
\flushbottom

\section{Introduction}

As the heaviest particle in the Standard Model (SM), the top quark plays an important role in both the precise test of the SM and  searches for new physics. Among the various production modes of a single top quark, the associated production of a top quark with a $W$ boson has the second-largest cross section at the Large Hadron Collider (LHC). The cross section for $tW$ production is directly proportional to the square of the Cabibbo-Kobayashi-Maskawa (CKM) matrix element $V_{tb}$. Inclusive and differential cross sections for $tW$ production have been measured extensively at the LHC~\cite{ATLAS:2012bqt,CMS:2012pxd,CMS:2014fut,ATLAS:2015igu,ATLAS:2016ofl,ATLAS:2017quy,CMS:2018amb,CMS:2022ytw,ATLAS:2024ppp,CMS:2024okz}, and all measurements agree well with theoretical predictions.

The $tW$ production  is particularly interesting due to its interference with the top quark pair production, which makes this process contain rich phenomenology in experiments~\cite{ATLAS:2018ivx}. 
Simultaneously, the interference leads to difficulties in making precise theoretical predictions. Substantial efforts have been dedicated to disentangling the $tW$ signals from $t\bar{t}$ production and addressing interference effects  at tree level~\cite{Belyaev:1998dn, Tait:1999cf, Campbell:2005bb, Frixione:2008yi, Demartin:2016axk}  and one-loop level~\cite{Dong:2024jxp}. 

The next-to-leading order (NLO) quantum chromodynamics (QCD) corrections to $tW$ production have been known for nearly three decades~\cite{Giele:1995kr, Cao:2008af, Kant:2014oha,Campbell:2005bb}.  
Additionally, the contributions from soft-gluon radiation have been explored through higher-order expansions~\cite{Kidonakis:2006bu,Kidonakis:2010ux,Kidonakis:2016sjf,Kidonakis:2021vob}, and an all-order resummation within the Soft-Collinear Effective Theory (SCET) framework has been performed~\cite{Li:2019dhg}. 
A combination of the NLO QCD calculation and the parton showers at NLO was provided by~\cite{Frixione:2008yi, Re:2010bp, Jezo:2016ujg}.

Progress toward next-to-next-to-leading order (NNLO) QCD corrections includes calculations of the two-loop hard function~\cite{Chen:2021gjv,Long:2021vse,Wang:2022enl,Chen:2022ntw,Chen:2022yni,Chen:2022pdw} and the two-loop $N$-jettiness soft function~\cite{Li:2016tvb,Li:2018tsq}. Despite these advances, a complete NNLO QCD correction for $tW$ production remains an open challenge.

In this paper, we present the calculation of the two-loop soft function for $tW$ production in the threshold limit, where the invariant mass of the $tW$ pair is close to the collider energy. In this limit, the cross section can be factorized as a convolution of the parton distribution functions (PDFs), the hard function and the soft function. 
The complete two-loop soft function is an indispensable ingredient for next-to-next-to-next-to-leading logarithmic resummation of the soft gluon effects, which can be used to improve the theoretical accuracy near the threshold region.
In the calculation, we make use of the reverse unitary method~\cite{Cutkosky:1960sp,Anastasiou:2002yz} to convert the phase space constraint into the form of propagators. The soft integrals are reduced to a set of master integrals by integration-by-part (IBP)~\cite{Tkachov:1981wb,Chetyrkin:1981qh} identities  using {\tt FIRE6}~\cite{Smirnov:2019qkx} and the master integrals are calculated with the method of differential equations~\cite{Kotikov:1990kg,Kotikov:1991pm,  Gehrmann:1999as,Henn:2013pwa}. The final results are given in terms of multiple polylogarithms (MPLs) \cite{2011arXiv1105.2076G}.

The rest of this paper is organized as follows.  In section~\ref{sec:factorized}, we discuss the factorization of the cross section near the threshold. The calculations of the NLO and NNLO soft functions are presented in section~\ref{sec:1loop} and section~\ref{sec:2loop}, respectively. 
In section~\ref{sec:expansion}, we explore the asymptotic behavior of the soft function in the low- and high-energy limits. 
Finally, we conclude in section~\ref{sec:conclu}.

\section{Factorization of the cross section near threshold}\label{sec:factorized}

The process we are considering is $b/\bar{b}\left( {{p_1}} \right) +  g\left( {{p_2}} \right) \to t/\bar{t}\left( {{p_3}} \right) + W^-/W^+\left( {{p_4}} \right)$ where $p_3$ and $p_4$ denote the momenta of the outgoing top/anti-top quark and $W$ boson, respectively,
which satisfy $p_3^2=m_t^2$ and $p_4^2=m_W^2$. 
We introduce the light-cone vectors $v_1$ and $v_2$
\begin{align}
    v_1^\mu  = \left( {1,0,0,1} \right),\quad v_2^\mu  = \left( {1,0,0, - 1} \right)\,,
\end{align}
which are along with the directions of $p_1$ and $p_2$, respectively.
The direction of the top quark momentum is expressed as
\begin{align}
    \quad v_3^\mu  = \left( {1,0,\beta \sin\theta ,\beta \cos\theta } \right)\,,
\end{align}
where $\beta$ is the velocity of the top quark and $\theta$  is the polar angle of the momentum of the top quark\footnote{Note that the $\theta$ angle in this soft function is different from that used in the hard function \cite{Chen:2022yni}.}. Explicitly, the momenta of colored particles in this process are given by 
\begin{align}
    p_1^\mu  = \frac{{\sqrt {{s_{12}}} }}{2} v_1^\mu ,\qquad p_2^\mu  = \frac{{\sqrt {{s_{12}}} }}{2} v_2^\mu ,\qquad p_3^\mu  = \frac{{{m_t}}}{{\sqrt {1-\beta^2} }}v_3^\mu\,,
\end{align}
with $s_{12}=(p_1+p_2)^2$.

In the threshold region, the invariant mass $Q$ of the $tW$ system is close to the hadronic energy of the collider, $\sqrt{S}$.
This region is described by the hadronic threshold variable $R\to 0$ with
\begin{align}
     R = 1 - \frac{Q^2}{S} = 1 - \frac{{{x_1}{x_2}{Q^2}}}{{{s_{12}}}}=\frac{{{s_{12}} - \left( {1 - \Delta {x_1}} \right)\left( {1 - \Delta {x_2}} \right){Q^2}}}{{{s_{12}}}}= \tau  + \Delta {x_1} + \Delta {x_2} \label{delta relation}\,,
\end{align}
where $x_i$ are the momentum fractions of the initial-state partons in the protons and $\Delta x_i=1-x_i$.
The partonic threshold variable $\tau$ is defined by
\begin{align}
    \tau  &= 1 - \frac{{{Q^2}}}{{{s_{12}}}} = \frac{{{{\left( {{p_t} + {p_W} + {p_\omega }} \right)}^2} - {{\left( {{p_t} + {p_W}} \right)}^2}}}{s_{12}} = \frac{2{E_\omega }}{\sqrt{s_{12}}} \equiv \frac{\omega }{\sqrt{{s_{12}}}} \,,
\end{align}
where $p_{\omega}^{\mu}$ denotes the momentum sum of all soft radiations and $E_{\omega}=p_{\omega}^0$ is the energy in the partonic center-of-mass frame.
It is clear that the hadronic threshold cannot be approached at any real colliders.
However, the partonic threshold region can still be enhanced because the parton distribution functions drop fast  at high $x_i$ \cite{Becher:2007ty}.
Therefore, the threshold contribution is dominant in the higher-order corrections as we show in ref. \cite{Li:2019dhg}.

In the hadronic threshold limit, the cross section takes a factorization form \cite{Li:2019dhg}, 
\begin{align}
 \frac{{d\sigma }}{dR d\Phi_2}  &=
   \int^1_0 {d{x_1}}  {f_1}\left( {{x_1},\mu } \right)\int^1_0d{x_2}{f_2}\left( {{x_2},\mu } \right)\frac{1}{2x_1 x_2 S}
   \times  H\left( {\mu ,\beta ,y} \right) 
   \nonumber\\ &
   \times \int^{\infty}_0 d\tau~s \left( {\tau ,\mu ,\beta ,y} \right)\delta \left( R-\tau  - \Delta {x_1} - \Delta {x_2}  \right)\, ,
   \label{eq:factorized} 
\end{align}
where $ d\Phi_2$ denotes the two-body phase space and $y\equiv \cos\theta$.
$f_i(x_i,\mu)$ represent the PDFs with the momentum fractions $x_i$ at the scale $\mu$.
$H$ and $s$ are the  hard function and soft function, which describe the interactions at the hard and soft scales, respectively. 
In momentum space the soft function is defined as the vacuum matrix element
\begin{align}
    s\left( {\tau , \mu, \beta ,y} \right) & =\sqrt{s_{12}} \int {\frac{{d{x_0}}}{{4\pi }}} {e^{\frac{{i \sqrt{s_{12}}\tau x_0 }}{2}}}\times  \nonumber \\
& \quad   \sum\limits_{{X_s}} {\left\langle {0\left| {\bar T\left\{ {Y_{{v_1}}^\dag Y_{{v_2}} {Y_{{v_3}}}} \right\}(x)} \right|{X_s}} \right\rangle \left\langle {{X_s}\left| {T\left\{ {{Y_{{v_1}}}{Y_{{v_2}}^{\dagger}}Y_{{v_3}}^\dag } \right\}} (0)\right|0} \right\rangle}
 \label{bare_soft}
\\
 &   = \sum\limits_{{X_s}} {\left\langle {0\left| {\bar T\left\{ {Y_{{v_1}}^\dag Y_{{v_2}} {Y_{{v_3}}}} \right\}(0)} \right|{X_s}} \right\rangle \delta \left( {\tau  -   {\frac{{{v_0}\cdot{\hat{p}_{{X_s}}}}}{{\sqrt {{s_{12}}} }}}} \right)\left\langle {{X_s}\left| {T\left\{ {{Y_{{v_1}}}{Y_{{v_2}}^{\dagger}}Y_{{v_3}}^\dag } \right\}} (0)\right|0} \right\rangle }
   \nonumber
\end{align}
with  $x=(x_0,0,0,0)$, $v_0=v_1+v_2$ and $\hat{p}_{X_s}$ extracting all the momenta of the final-state soft particles. $T(\bar T)$ is the (anti-)time-ordering operator, and $Y_{{v_1}}, Y_{{v_2}}^\dag$ and $Y_{{v_3}}^\dag$ are the soft Wilson lines defined by \cite{Beneke:2002ph,Chay:2004zn,Korchemsky:1991zp}
\begin{align}
    {Y_{{v_1}}}\left( x \right) &= P\exp \left( {-i{g_s}\int\limits_{ - \infty }^0 {ds} {v_1} \cdot A_s^a\left( {x + s{v_1}} \right)\mathbf{T}_1^a} \right)\,,\\
    {Y^\dag_{{v_2}}}\left( x \right) &= {\bar P}\exp \left( {-i{g_s}\int\limits_{ - \infty }^0 {ds} {v_2} \cdot A_s^a\left( {x + s{v_2}} \right)\mathbf{T}_2^a} \right)\,,\\
    Y_{{v_3}}^\dag \left( x \right) 
    &= P\exp \left( {i{g_s}\int\limits_0^{ + \infty } {ds} {v_3} \cdot A_s^a\left( {x + s{v_3}} \right)\mathbf{T}_3^a} \right)\,,
\end{align}
where $P$ ($\bar P$) is the (anti-)path-ordering operator\footnote{The (anti-)path-ordering operator rearranges the fields in such a way that the fields with higher values of $s$ are to the left (right).  } and  $\mathbf{T}_i$ represents the color charge associating with the $i$-th parton~\cite{Catani:1996jh,Catani:1996vz}~\footnote{Because we have used the color charge, the explicit forms of the soft Wilson lines may be different from the others used in previous literature.}. 
For the initial- and final-state quark, 
the color-charge matrix is given by $\mathbf{T}^a_{ij} = -t^a_{ji}$ and 
 $\mathbf{T}^a_{ij} = t^a_{ij}$, respectively.
 For the gluon, the color-charge matrix is $\mathbf{T}^a_{bc} = -i f_{abc}$. 
 Note that the color charge operators along different directions commutate with each other, such as $[\mathbf{T}_1^a,\mathbf{T}_2^b]=0$.
 As a result, the relative ordering of the Wilson lines $Y_{{v_1}}, Y_{{v_2}}^\dag$ and $Y_{{v_3}}^\dag$ in eq. \eqref{bare_soft} can be arbitrary.

It is convenient to perform the Laplace transformation
\begin{align}
     \frac{d\tilde{\sigma}(t,\mu)}{d\Phi_2} =&\int\limits_0^\infty {dR}~\frac{{d\sigma }}{{dR d\Phi_2}}~\exp \left( { - \frac{R}{{t}}} \right)\,,
     \nonumber \\
    {{\tilde f}_i}\left( t,\mu   \right) =& \int\limits_0^1 {d{x_i}}~{f_i}\left( {{x_i},\mu } \right)~\exp\left( -\frac{1-x_i}{t} \right)\,,
    \nonumber \\ 
    \tilde{s}(t,\mu,\beta, y) = & \int_0^{\infty} d\tau~ s(\tau, \mu, \beta, y)~\exp\left( -\frac{\tau}{t}\right)\,,
\end{align}
which converts the cross section to a product of the PDFs, hard function and soft function,
\begin{align}~\label{eq:crossLap}
    \frac{d\tilde{\sigma}(t,\mu)}{d\Phi_2} 
    &=\frac{1}{2s_{12}} H\left( \mu ,\beta ,y  \right){{\tilde f}_1}\left( {t,\mu } \right){{\tilde f}_2}\left( {t,\mu } \right)\tilde s\left( {t,\mu ,\beta ,y } \right)\,.
\end{align}

Because the cross section is scale invariant, the renormalization group (RG) evolutions of the various components in eq.~(\ref{eq:crossLap}) satisfy the identity below order-by-order in the strong coupling $\alpha_s$, 
\begin{align}
    \frac{d\ln H\left( \mu ,\beta ,y  \right)}{d\ln\mu} + \frac{d \ln{\tilde f}_1(t,\mu)}{d\ln\mu}+ \frac{d\ln {\tilde f}_2 \left( {t,\mu } \right)}{d\ln\mu} + \frac{d\ln\tilde s\left( {t,\mu ,\beta ,y } \right)}{d\ln\mu}~= 0\,. 
\end{align}
Therefore, the soft anomalous dimension can be derived from the independence of the cross section on the renormalization scale $\mu$,
\begin{align}
    {\gamma _s} = \frac{{d\ln \tilde s}}{{d\ln \mu }} =  - \frac{{d\ln H}}{{d\ln \mu }} - \frac{{d\ln {{\tilde f}_1}}}{{d\ln \mu }} - \frac{{d\ln {{\tilde f}_2}}}{{d\ln \mu }}\,.
    \label{eq:gammas}
\end{align}
The RG equation of the hard function can be obtained from the universal anomalous dimensions; see ref.~\cite{Ferroglia:2009ii}.
The RG evolution of the PDFs in the threshold limit  can be found in refs.~\cite{Korchemsky:1992xv,Moch:2004pa}.  
According to eq.~(\ref{eq:gammas}), the anomalous dimension of the soft function is given by
\begin{align}
    {\gamma _s} =  &- \left( {{\mathbf{T}}_1^2 + \mathbf{T}_2^2} \right){\gamma _{{\rm{cusp}}}}\left( {\ln {t ^2} + \ln \frac{{{s_{12}}}}{{{\mu ^2}}}} \right) + \mathbf{T} _1\cdot \mathbf{T}_3{\gamma _{{\rm{cusp}}}}\ln \frac{{{{\left( {1 - \beta y} \right)}^2}}}{{1 - {\beta ^2}}}\nonumber\\
    &+ \mathbf{T}_2 \cdot \mathbf{T}_3{\gamma _{{\rm{cusp}}}}\ln \frac{{{{\left( {1 + \beta y} \right)}^2}}}{{1 - {\beta ^2}}}- 2{\gamma ^q} - 2{\gamma ^g} - 2{\gamma ^Q} - {\rm{2}}\gamma _f^q - {\rm{2}}\gamma _f^g \,,
\end{align}
where all anomalous dimensions can be found in ref.~\cite{Ahrens:2010zv}. The renormalized soft function reads
\begin{align}
    \tilde s\left( {t,\mu ,\beta ,y} \right) =&  Z_s^{ - 1} \tilde s_{\rm bare}\left( {t,\mu ,\beta ,y } \right)\,.
  \label{eq:renormalization_constant}
\end{align}
Because the bare soft function is scale independent, the renormalization constant $Z_s$ satisfies the evolution equation 
\begin{align}
   \frac{d\ln Z_s}{d\ln \mu}= - \gamma_s \,.
\end{align}
Solving this equation in the $\overline{\rm MS}$ scheme, we obtain 
\begin{align}
    \ln {Z_s} =& \frac{{{\alpha _s}}}{{4\pi }}\left( {\frac{{\gamma^\prime {{_s^{\left( 0 \right)}} }}}{{4{\epsilon ^2}}} + \frac{{\gamma _s^{\left( 0 \right)}}}{{2\epsilon }}} \right)    
    +{\left( {\frac{{{\alpha _s}}}{{4\pi }}} \right)^2}\left( { - \frac{{3{\beta _0}\gamma^\prime {{_s^{\left( 0 \right)}} }}}{{16{\epsilon ^3}}} + \frac{{\gamma^\prime {{_s^{\left( 1 \right)}} } - 4{\beta _0}\gamma _s^{\left( 0 \right)}}}{{16{\epsilon ^2}}} + \frac{{\gamma _s^{\left( 1 \right)}}}{{4\epsilon }}}\right)  + \mathcal{O}(\alpha^3_s) \,,
\end{align}
where we have worked in $d=4-2\epsilon$ dimensional spacetime to regulate the divergences and adopted the perturbative expansion in $\alpha_s$,
\begin{align}
    \gamma_s =\sum_{i=0} \left( \frac{\alpha_s}{4\pi} \right)^{i+1}\gamma_s^{(i)}\,.
\end{align}
We have also used the notation $\gamma'_s= \partial \gamma_s /\partial\ln \mu $ and $\beta_0 = 11\mathrm{C_A}/3 - 2n_f/3$ with $\mathrm{C_A}=3$ and $n_f$ being the number of active quark flavors. 

After expanding eq. \eqref{eq:renormalization_constant} order-by-order in $\alpha_s/4\pi$, the renormalized NLO and NNLO soft functions in Laplace space are given by 
\begin{align}
    {{\tilde s}^{\left( 1 \right)}} &= {{\tilde s}_{\rm bare}^{\left( 1 \right)}} - Z_s^{\left( 1 \right)}\label{NLO ren}\,,\\
    {{\tilde s}^{\left( 2 \right)}} &= {{\tilde s}_{\rm bare}^{\left( 2 \right)}} - Z_s^{\left( 2 \right)} - {{\tilde s}_{\rm bare}^{\left( 1 \right)}}Z_s^{\left( 1 \right)} + \left(Z_s^{\left( 1 \right)}\right)^2 - \frac{{{\beta _0}}}{\epsilon }{{\tilde s}_{\rm bare}^{\left( 1 \right)}}\label{NNLO ren}\,.
\end{align}
All the divergences in the bare soft function $\tilde s_{\rm bare}$ are related to the renormalization constant $Z_s$ and thus the anomalous dimensions according to the above discussions. 

\section{NLO soft function}\label{sec:1loop}

The NLO bare soft function can be obtained by calculating the integral 
\begin{align}~\label{eq:s1Int}
    {s_{\rm bare}^{\left( 1 \right)}}\left( \tau,\mu,\beta,y  \right) = -\frac{{2{e^{{\gamma _E}\epsilon }}{\mu ^{2\epsilon }}}}{{{\pi ^{1 - \epsilon }}}}\int {{d^d}k} \delta \left( {{k^2}} \right)\theta \left( {{k^ 0 }} \right)J^{\mu \left( 0 \right)\dag }_a\left( k \right)J_{a \mu }^{ \left( 0 \right)}\left( k \right)F\left( {v_1,v_2,k} \right)\, ,
\end{align}
where the leading order soft current is defined as
\begin{align}
    J_a^{\mu \left( 0 \right)}\left( k \right) =- \sum\limits_{i = 1}^3 {\mathbf{T}^a_i\frac{{v_i^\mu }}{{{v_i} \cdot k}}}\,.
\end{align}
A prefactor $e^{{\gamma _E}\epsilon } (4\pi)^{-\epsilon}$ has been inserted due to our choice of the $\overline{\rm MS}$ scheme in renormalization of $\alpha_s$.
In eq.~(\ref{eq:s1Int}),  $F(v_1,v_2,k)$ is the measurement function in the threshold limit, defined as
\begin{align}
    F\left( {v_1,v_2,k} \right) = \delta \left( {\tau  - \frac{1}{{\sqrt {{s_{12}}} }}\left( {v_1 + v_2} \right) \cdot k} \right) = \sqrt {{s_{12}}} \delta \left( {\omega  - {v_0} \cdot k} \right)\,.
\end{align}
With the Lorentz indices contracted, the NLO soft function is represented by 
\begin{align}
    {s_{\rm bare}^{\left( 1 \right)}}\left( \tau,\mu, \beta,y  \right) & =- \frac{{2{e^{{\gamma _E}\epsilon }}{\mu ^{2\epsilon }}}}{{{\pi ^{1 - \epsilon }}}}\sqrt {{s_{12}}} \int {d^d}k~\delta\left( {{k^2}} \right)~\theta(k^0)~\delta \left( {\omega  - {v_0} \cdot k} \right)\nonumber\\
    &\quad \times \left[ {\frac{{2\mathbf{T}_1\cdot\mathbf{T}_2{v_1} \cdot {v_2}}}{{\left( {{v_1} \cdot k} \right)\left( {{v_2} \cdot k} \right)}} + \frac{{2\mathbf{T}_1\cdot\mathbf{T}_3{v_1} \cdot {v_3}}}{{\left( {{v_1} \cdot k} \right)\left( {{v_3} \cdot k} \right)}} + \frac{{2\mathbf{T}_2\cdot\mathbf{T}_3{v_2} \cdot {v_3}}}{{\left( {{v_2} \cdot k} \right)\left( {{v_3} \cdot k} \right)}} + \frac{{\mathbf{T}_3\cdot\mathbf{T}_3{v_3} \cdot {v_3}}}{{{{\left( {{v_3} \cdot k} \right)}^2}}}} \right]\nonumber\\
    & =\tau^{-1-2\epsilon} \left(\frac{\mu^2}{s_{12}}\right)^{\epsilon} s_{\rm bare}^{(1)}(\beta,y)\,.
    \label{eq:nlointegral}
\end{align}
In the third line, we have written the dependence on $\tau$ and $\mu$ explicitly in the result.
The color charges would act on the color bases of the hard function.
Due to its simple color configuration, there is only one  color basis up to all orders for the $bg\to tW$ process.
Therefore we can make the substitution in the above equation:
\begin{align}
    \mathbf{T}_1\cdot\mathbf{T}_2 \to  -\frac{\mathrm{C_A}}{2}, \quad
    \mathbf{T}_1\cdot\mathbf{T}_3 \to \frac{\mathrm{C_A}}{2}-\mathrm{C_F}, \quad
    \mathbf{T}_2\cdot\mathbf{T}_3 \to -\frac{\mathrm{C_A}}{2}, \quad 
    \mathbf{T}_3\cdot\mathbf{T}_3 \to \mathrm{C_F}
\end{align}
with $\mathrm{C_F}=4/3$ and $\mathrm{C_A}=3$ being the Casimir operators of $SU(3)$ in the fundamental and adjoint representations, respectively.

The first term in the integrand of eq.~(\ref{eq:nlointegral}) depends only on $v_1$ and $v_2$, and thus is easy to compute.
The other terms depend on $v_3$ and deserve a detailed discussion.
They can be represented by the integral family 
\begin{align}
    {F_{{a_1},{a_2}}} = \int {\left[ {dk} \right]\delta \left( {\omega  - {v_0} \cdot k} \right)} \frac{1}{{{{\left( {{v_1} \cdot k} \right)}^{{a_1}}}{{\left( {{v_3} \cdot k} \right)}^{{a_2}}}}}
\end{align}
with $[dk]\equiv d^d k \delta(k^2)\theta(k^0)$.
The result for the integral containing $v_2$ can be obtained by replacing $y$ by $-y$.
We treat the $\delta$-functions  by making use of the reverse unitary method to change them in the form of propagators, i.e.,
\begin{align}
   \delta\left( {{k^2}} \right)\theta(k^0) &= \frac{1}{{2\pi i}}\left( {\frac{1}{{{k^2} - i0^+}} - \frac{1}{{{k^2} + i0^+}}} \right)\theta(k^0)\,,
   \label{eq:deltafun}\\
    \delta \left( {\omega  - {v_0} \cdot k} \right) &= \frac{1}{{2\pi i}}\left( {\frac{1}{{\omega  - {v_0} \cdot k - i0^+}} - \frac{1}{{\omega  - {v_0} \cdot k + i0^+}}} \right)\,.
\end{align}
As a result, the integrals in this family can be reduced to a finite set of master integrals because of the relations among them derived from the IBP identities~\footnote{The $\theta(k^0)$ term in eq. (\ref{eq:deltafun}) does not affect the IBP identities because its derivative gives $\delta(k^0)$, which causes scaleless integrals.  }.
We have employed the package {\tt FIRE6}, which implements the algorithms proposed in ref. \cite{Laporta:2000dsw}, to perform the reduction, and found the following master integrals,
\begin{align}
    \vec f\left( {\omega ,\beta ,y,\epsilon } \right) = {\left( {{F_{0,0}},{F_{0,1}},{F_{1,1}}} \right)^T}\,.
\end{align}

Then we calculate the master integrals analytically with the method of differential equations.
In particular, we have transformed the differential equation into the canonical form  \cite{Henn:2013pwa}
\begin{align}
    \frac{\partial  \vec g\left( {\beta ,y,\epsilon} \right) }{\partial \beta}= \epsilon \hat B\left( {\beta ,y} \right)\vec g\left( {\beta ,y,\epsilon} \right)\label{NLO differential equation}
\end{align}
by choosing a proper basis $\vec g\left( {\beta ,y,\epsilon} \right) = \hat T\left( {\omega ,\beta ,y,\epsilon } \right)\vec f\left( {\omega ,\beta ,y,\epsilon } \right)$
with 
\begin{align}
    \hat T\left( {\omega ,\beta ,y,\epsilon } \right) = \frac{{2\Gamma \left( {1 - 2\epsilon } \right)}}{{{\omega ^{1 - 2\epsilon }}{\pi ^{1 - \epsilon }}\Gamma \left( {1 - \epsilon } \right)}}\left( {\begin{array}{*{20}{c}}{1-2\epsilon }&0&0\\0&{\epsilon \omega \beta }&0\\0&0&{\epsilon {\omega ^2}\left( {1 - \beta y} \right)}\end{array}} \right)\,.
\end{align}
Note that the new basis $\vec g$ is dimensionless and thus does not depend on $\omega$.
The coefficient matrix $\hat B$ is independent of the dimensional regulator, and can be expressed by
\begin{align}
    \hat B\left( {\beta ,y} \right) = \frac{{\hat a}}{{\beta  - 1}} + \frac{{\hat b}}{\beta } + \frac{{\hat c}}{{\beta  + 1}} + \frac{{\hat d}}{{\beta  - 1/y}}
\end{align}
with 
\begin{align}
    \hat a = \left( {\begin{array}{*{20}{c}}0&0&0\\{ - 1}&{ - 1}&0\\{ - 2}&{ - 2}&0\end{array}}\right),
   ~~ \hat b = \left( {\begin{array}{*{20}{c}}0&0&0\\0&2&0\\4&0&2 \end{array}} \right),
   ~~ \hat c = \left( {\begin{array}{*{20}{c}}0&0&0\\1&{ - 1}&0\\{ - 2}&2&0\end{array}} \right),
    ~~ \hat d = \left( {\begin{array}{*{20}{c}}0&0&0\\0&0&0\\0&0&{ - 2}\end{array}} \right)\,.
\end{align}
We are interested in the result of $\vec g$ as an expansion in a series of $\epsilon$,
\begin{align}
    \vec g\left( {\beta ,y,\epsilon} \right) = \sum\limits_{n = 0}^\infty  {{{\vec g}^{\left( n \right)}}\left( {\beta ,y} \right){\epsilon ^n}}\,,
\end{align}
where we have chosen  deliberately  the normalization factors of the basis integrals so that no negative powers of $\epsilon$ appear.
Then we can solve the canonical differential equation order-by-order in $\epsilon$,
\begin{align}
    {{\vec g}^{\left( {n + 1} \right)}}\left( {\beta ,y} \right) = \int\limits_{{0}}^\beta  {d\beta^\prime } \hat B\left( {\beta^\prime ,y} \right){{\vec g}^{\left( n \right)}}\left( {\beta^\prime ,y} \right) + {{\vec g}^{\left( {n + 1} \right)}}\left( {{0},y} \right)\,, 
    \label{canonial solution}
\end{align}
where ${{\vec g}^{\left( {n} \right)}}\left( {{0},y} \right)$ is the boundary value. 

Because there is no new divergence if setting the top quark at rest, the master integrals are regular at the phase space point of $\beta=0$.
Given that the transformation matrix $\hat T$ is also regular, we derive the following regularity condition
\begin{align}
    0 = \mathop {\lim }\limits_{\beta  \to 0} \beta {\partial _\beta }\vec g\left( {\beta ,y,\epsilon} \right) = \mathop {\lim }\limits_{\beta  \to 0} \beta \hat B\left( {\beta ,y} \right)\vec g\left( {\beta ,y,\epsilon} \right)\,,
\end{align}
which leads to
\begin{align}
    \vec g\left( {0,y,\epsilon} \right) = {\left( {{g_1},0, - 2{g_1}} \right)^T}\,.
\end{align}
Therefore, we only need to calculate the first basis integral $g_1$, which is simple, 
\begin{align}
    {g_1}\left( {0,y,\epsilon} \right) = \frac{{2\Gamma \left( {2 - 2\epsilon } \right)}}{{{\omega ^{1 - 2\epsilon }}{\pi ^{1 - \epsilon }}\Gamma \left( {1 - \epsilon } \right)}}\int {\left[ {dk} \right]\delta \left( {\omega  - {v_0} \cdot k} \right)}=1\,.
\end{align}
Another advantage of choosing the boundary condition in this way is that the master integrals at $\beta=0$ do not depend on $y$.
As a result, we do not have to construct and solve the differential equation with respect to $y$.

The full result of the basis integrals 
can be obtained using eq.~(\ref{canonial solution}),
and  can be written in terms of MPLs, 
which are defined by $G(\beta)=1$ and
\begin{align}
    G\left( {{a_1}, \ldots ,{a_n};\beta } \right) &= \int\limits_0^\beta  {\frac{{d t }}{{t  - {a_1}}}} G\left( {{a_2}, \ldots ,{a_n};t } \right)\,,\\
    G(\vec 0_n;\beta)&=\frac{1}{n!}\ln^n\beta\,.
\end{align}
The MPLs have nice properties~\cite{Duhr:2014woa} and can be evaluated using the packages {\tt GiNaC} \cite{Bauer:2000cp} and {\tt PolyLogTools}~\cite{Duhr:2019tlz}.
We provide the analytic results for all the basis integrals in an auxiliary file.

The NLO bare soft function in the Laplace space  is
\begin{align}
\tilde s_{\rm bare}^{(1)}\left( {t,\mu ,\beta ,y} \right) = \Gamma \left( { - 2\epsilon } \right)
\left( \frac{\mu^2} {s_{12}t^2}\right)^{\epsilon} 
s_{\rm bare}^{(1)}(\beta,y)\,.
\end{align}
Performing renormalization according to eq.~(\ref{NLO ren}), we obtain the NLO renormalized soft function,
\begin{align}
  &{{\tilde s}^{\left( 1 \right)}}\left( {L,\beta,y} \right)= L^2 \mathrm{(C_A +C_F)}+L\big[2\mathrm{C_A}\left(G_{-1/y}-G_{1/y}\right)+2\mathrm{C_F}\left(-1-G_{-1}-G_{1}+2G_{1/y}\right)\big]\nonumber\\
  &\frac{\mathrm{C_A}}{6}\left(\pi^2-24G_{0,-1/y}+24G_{0,1/y}-12G_{-1/y,-1}-12G_{-1/y,1}+24G_{-1/y,-1/y}+12G_{1/y,-1}\right.\nonumber\\
  &\left.+12G_{1/y,1}-24G_{1/y,1/y}\right)+\frac{\mathrm{C_F}}{6}\bigg[\frac{12}{\beta}\left(G_{-1}-G_{1}\right)+\pi^2-12G_{-1,-1}+12G_{-1,1}+24G_{0,-1}\nonumber\\
  &+24G_{0,1}-48G_{0,1/y}+12G_{1,-1}-12G_{1,1}-24G_{1/y,-1}-24G_{1/y,1}+48G_{1/y,1/y}\bigg]\, ,
\end{align}
where $ L = \ln \frac{{{s_{12}}{t^2}}}{{\mu ^2}}$ and $G_{a_1,\ldots,a_n}\equiv G\left( {{a_1}, \ldots ,{a_n};\beta } \right)$.

\section{NNLO soft function}\label{sec:2loop}

The NNLO bare soft function consists of two parts, i.e., the double-real and virtual-real corrections, 
and thus can be expressed as
\begin{align}
    s_{\mathrm{bare}}^{\left( 2 \right)}\left( \tau,\mu,\beta,y  \right) & = 
    s_{\mathrm{DR}}^{\left( 2 \right)}\left( \tau,\mu,\beta,y  \right) 
    + s_{\mathrm{VR}}^{\left( 2 \right)}\left( \tau,\mu,\beta,y  \right)\nonumber\\
    & =\tau^{-1-4\epsilon} \left(\frac{\mu^2}{s_{12}}\right)^{2\epsilon} s_{\rm bare}^{(2)}(\beta,y)\,.
\end{align}

\begin{figure}
    \centering
    \includegraphics[width=0.7\linewidth]{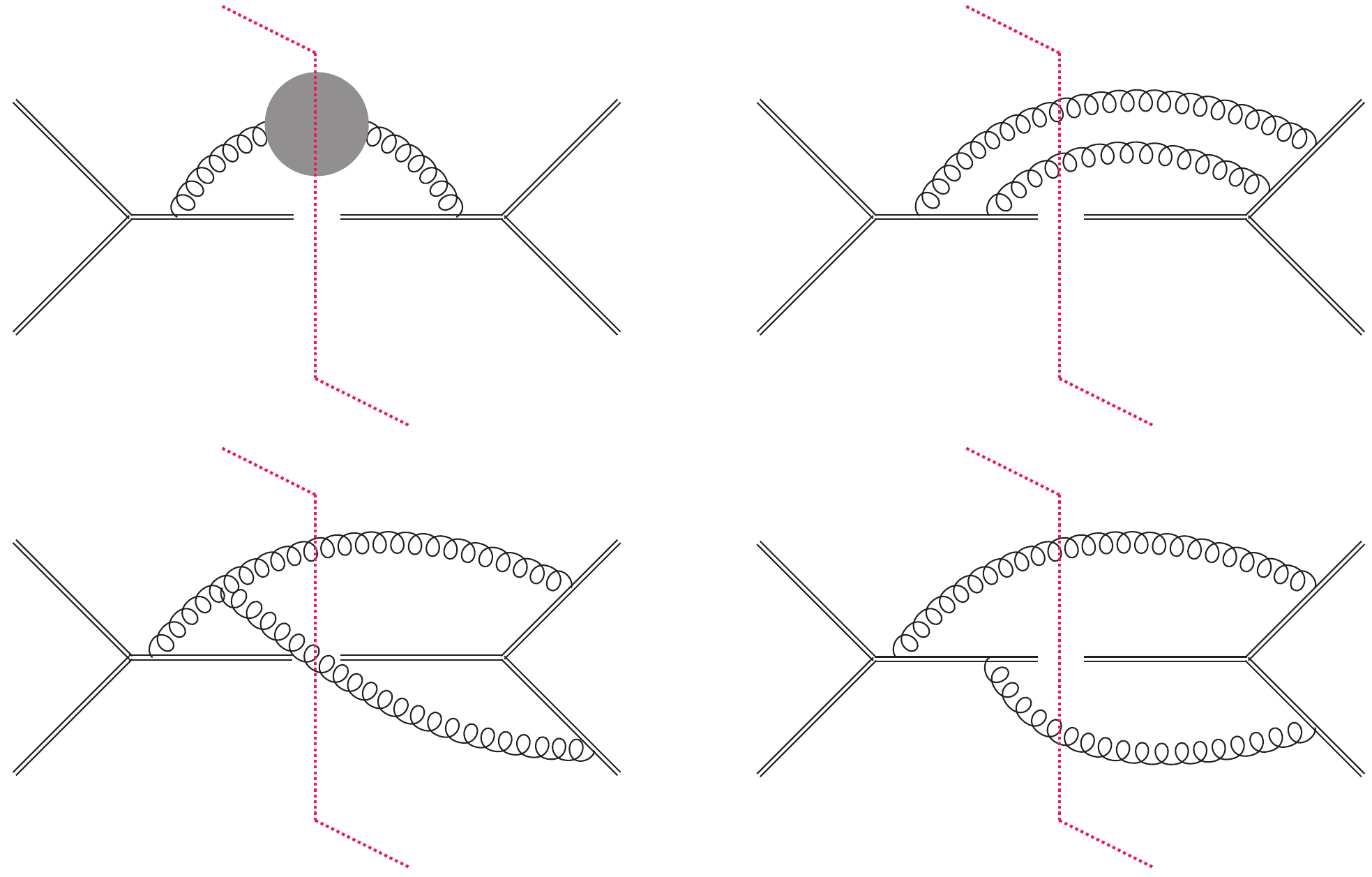}
    \caption{Sample double-real diagrams contributing to the NNLO soft function.
    The blob in the first diagram denotes the  gluon self-energy subgraph. }
    \label{fig:DR}
\end{figure}

The selected Feynman diagrams for the double-real corrections are shown in figure~\ref{fig:DR}. 
Depending on the final state, this part splits further into two terms,
\begin{align}
    s^{\left( 2 \right)}_{\mathrm{DR}}\left( \tau,\mu,\beta,y  \right)=s^{\left( 2 \right)}_{gg}\left( \tau,\mu,\beta,y  \right)+s^{\left( 2 \right)}_{q \bar q}\left( \tau,\mu,\beta,y  \right)\,.
    \label{eq:sdr}
\end{align}
The emission of two soft gluons contributes to the first term, 
\begin{align}
    {s^{\left( 2 \right)}_{gg}}\left( \tau,\mu,\beta,y  \right) = &\frac{{2{e^{2{\gamma _E}\epsilon }}{\mu ^{4\epsilon }}}}{{{\pi ^{2 - 2\epsilon }}}}\int {{d^d}{k_1}} d^d{k_2}\delta \left( {k_1^2} \right)\theta \left( {k_1^ 0 } \right)\delta \left( {k_2^2} \right)\theta \left( {k_2^ 0 } \right)\\
    &\times J_{{a_1}{a_2}}^{{\mu _1}{\nu _1}\left( 0 \right)\dag }\left( {{k_1},{k_2}} \right){d_{{\mu _1}{\mu _2}}}\left( {{k_1}} \right){d_{{\nu _1}{\nu _2}}}\left( {{k_2}} \right)J_{{a_1}{a_2}}^{{\mu _2}{\nu _2}\left( 0 \right)}\left( {{k_1},{k_2}} \right)F\left( {v_1,v_2,{k_1},{k_2}} \right)\,,\nonumber
\end{align}
where the two-gluon soft current is given by \cite{Catani:1999ss,Czakon:2011ve}
\begin{align}
    J_{{a_1}{a_2}}^{\mu \nu \left( 0 \right)}\left( {{k_1},{k_2}} \right) = &\frac{1}{2}\left\{ {J_{{a_1}}^{\mu \left( 0 \right)},J_{{a_2}}^{\nu \left( 0 \right)}} \right\} + i{f_{{a_1}{a_2}{a_3}}}\sum\limits_{i = 1}^3 \mathbf{T}_i^{a_3}\\
    &\times \left[ {\frac{{v_i^\mu k_1^\nu  - v_i^\nu k_2^\mu }}{{{k_1} \cdot {k_2}{v_i} \cdot \left( {{k_1} + {k_2}} \right)}} - \frac{{{v_i} \cdot \left( {{k_1} - {k_2}} \right)}}{{2{v_i} \cdot \left( {{k_1} + {k_2}} \right)}}\left( {\frac{{v_i^\mu v_i^\nu }}{{{v_i} \cdot {k_1}{v_i} \cdot {k_2}}} + \frac{{{g^{\mu \nu }}}}{{{k_1} \cdot {k_2}}}} \right)} \right]\,.\nonumber
\end{align}
Because of current conservation, one can take  the polarization tensor  $d_{{\mu _1}{\mu _2}}=-g_{\mu_1 \mu_2}$ for simplicity.
The second term in eq.~(\ref{eq:sdr}) represents the effect of a soft quark-antiquark pair,
\begin{align}
    {s^{\left( 2 \right)}_{q\bar q}}\left( \tau,\mu,\beta,y  \right) = &\frac{{2{e^{2{\gamma _E}\epsilon }}{\mu ^{4\epsilon }}}}{{{\pi ^{2 - 2\epsilon }}}}\int  [d {k_1}] [d {k_2}] \sum\limits_{i,j = 1}^3 {{\mathcal{T}_{ij}}\left( {{k_1},{k_2}} \right)F\left( {v_1,v_2,{k_1},{k_2}} \right)}
\end{align}
with
\begin{align}
    {\mathcal{T}_{ij}}\left( {{k_1},{k_2}} \right)=- \mathrm{T_F}\mathbf{T}_i \cdot \mathbf{T}_j\frac{{2{v_i} \cdot {v_j}{k_1} \cdot {k_2} + {v_i} \cdot \left( {{k_1} - {k_2}} \right){v_j} \cdot \left( {{k_1} - {k_2}} \right)}}{{2{{\left( {{k_1} \cdot {k_2}} \right)}^2}{v_i} \cdot \left( {{k_1} + {k_2}} \right){v_j} \cdot \left( {{k_1} + {k_2}} \right)}}\,,
\end{align}
which can be found in \cite{Catani:1999ss}. 
We have checked that these expressions can be reproduced from the expansion of the soft Wilson lines.
In the case of double emissions, the measurement function
 $F\left( {v_1,v_2,{k_1},{k_2}} \right)$ is defined as
\begin{align}
    F\left( {v_1,v_2,{k_1},{k_2}} \right)=\sqrt {{s_{12}}} \delta \left( {\omega  - {v_0} \cdot (k_1+k_2)} \right)\,.
\end{align}

\begin{figure}[t]
    \centering
    \includegraphics[width=0.9\linewidth]{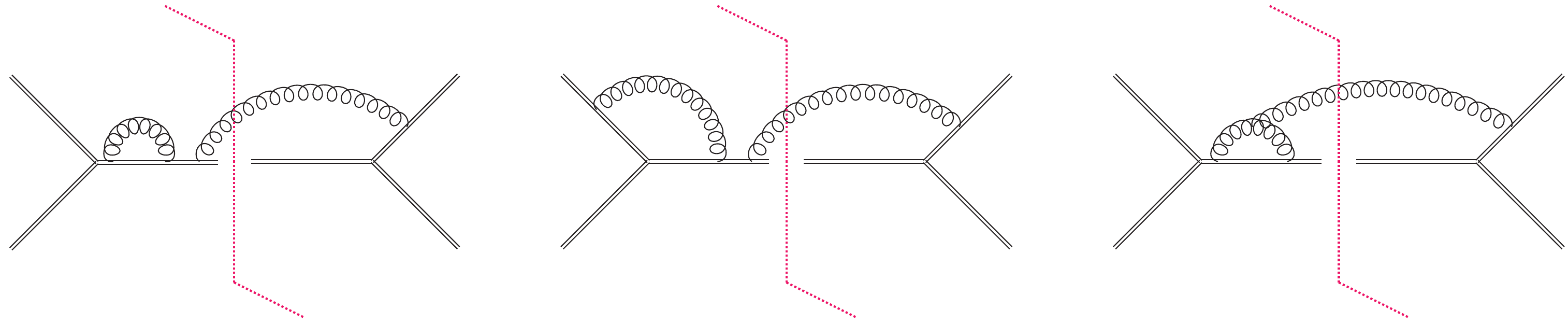}
    \caption{Sample virtual-real diagrams contributing to the NNLO soft function.}
    \label{fig:2}
\end{figure}

Figure~\ref{fig:2} shows  selected Feynman diagrams relevant to virtual-real corrections. 
Their contributions to the NNLO soft function can be written as
\begin{align}
    s_{\mathrm{VR}}^{\left( 2 \right)}\left( \tau,\mu,\beta,y  \right) = \frac{{4{e^{2{\gamma _E}\epsilon }}{\mu ^{4\epsilon }}}}{{{\pi ^{1 - \epsilon }}}}{\mathop{\rm Re}\nolimits} \left[ {\int { [dk] J^{\mu \left( 0 \right)\dag }_a\left( k \right)J_{a \mu }^{ \left( 1 \right)}\left( k \right) F\left( {v_1,v_2,k} \right)} } \right]\,.
\end{align}
The analytic result of the one-loop soft current $J^{\nu \left( 1 \right)}_a\left( k \right)$ can be found in~\cite{Bierenbaum:2011gg}. 
However, it is not suitable for performing the integration over $k$.
Therefore, we use the form of $J^{\nu \left( 1 \right)}_a\left( k \right)$ with the soft loop momentum $l$ unintegrated.

The scalar integrals needed in the calculation of double-real corrections can be written as 
\begin{align}
    F_{{a_1},{a_2},{a_3},{a_4},{a_5},{a_6}}^{\left( i \right)} = \int {\left[ {d{k_1}} \right]} \left[ {d{k_2}} \right]\delta \left( {\omega  - {v_0} \cdot \left( {{k_1} + {k_2}} \right)} \right)\prod\limits_{j = 1}^6 {\frac{1}{{{{\left( {{D_j^{(i)}}} \right)}^{{a_j}}}}}}\, ,
\end{align}
where $D_j^{(i)}$ is the $j$-th denominator in the $i$-th family.
After performing reduction, 
we find two integral families with the following denominators:
\begin{align}
    &D_j^{(1)}: \quad {v_1} \cdot {k_2},~{v_1} \cdot \left( {{k_1} + {k_2}} \right),~{v_2} \cdot {k_1},~{v_3} \cdot {k_1},~{v_3} \cdot \left( {{k_1} + {k_2}} \right),~{\left( {{k_1} + {k_2}} \right)^2} \,, \\
    &D_j^{(2)}: \quad {v_1} \cdot {k_1},~{v_1} \cdot {k_2},~{v_2} \cdot {k_1},~{v_3} \cdot {k_1},~{v_3} \cdot {k_2},~{\left( {{k_1} + {k_2}} \right)^2}\,.
\end{align}
There are $13$ and $12$ master integrals for the first and second families, respectively,
which can be chosen as
\begin{align}
    {{\vec f}^{(1)}}&\left( {\omega ,\beta ,y,\epsilon } \right) = ({F^{(1)}_{0,0,0,0,0,0}},{F^{(1)}_{0,0,0,1,0,0}},{F^{(1)}_{0,0,0,0,1,0}},{F^{(1)}_{0,0,0,1,1,0}},{F^{(1)}_{1,0,0,0,1,0}},{F^{(1)}_{1,0,0,1,1,0}},\nonumber\\
    &{F^{(1)}_{1,0,0,1,0,1}},{F^{(1)}_{1,0,0,1, - 1,1}},{F^{(1)}_{0,1,0,1,0,0}},{F^{(1)}_{0,1,0,0,1,0}},{F^{(1)}_{ - 1,1,0,1,0,0}},{F^{(1)}_{1, - 1,0,1,0,1}},{F^{(1)}_{1,1,0,1,1,0}})^T\,,\nonumber
\end{align}
and
\begin{align}
    {{\vec f}^{\left( 2 \right)}}\left( {\omega ,\beta ,y,\epsilon } \right) =(&{F^{(2)}_{0,0,0,0,0,0}},{F^{(2)}_{0,0,0,1,0,0}},{F^{(2)}_{0,0,1,1,0,0}},{F^{(2)}_{1,0,0,1,0,0}},{F^{(2)}_{0,0,0,0,1,0}},{F^{(2)}_{0,1,0,0,1,0}},\nonumber\\
    &{F^{(2)}_{0,0,0,1,1,0}},{F^{(2)}_{1,0,0,1,1,0}},{F^{(2)}_{0,1,0,1,1,0}},{F^{(2)}_{0,0,1,1,1,0}},{F^{(2)}_{0,1,1,1,1,0}},{F^{(2)}_{1,1,0,1,1,0}})^T\,.
\end{align}

The scalar integrals for virtual-real corrections can be written as 
\begin{align}
    F_{{a_1},{a_2},{a_3},{a_4},{a_5},{a_6},{a_7}}^{\left( 3 \right)} = \int {\left[ {d{k}} \right]} d^d l \delta \left( \omega  - {v_0} \cdot  {k} \right) \prod\limits_{j = 1}^7 {\frac{1}{{{{\left( {{D_j^{(3)}}} \right)}^{{a_j}}}}}}
\end{align}
with the denominators
\begin{align}
    D_j^{(3)}:\quad {v_1} \cdot k, ~ {v_1} \cdot l, ~ {v_2} \cdot \left( {k - l} \right), ~{v_3} \cdot k, ~{v_3} \cdot \left( {k - l} \right),~{l^2},~{\left( {k - l} \right)^2}\,.
\end{align}
All the scalar integrals are reduced to 9 master integrals,
\begin{align}
    {{\vec f}^{\left( 3 \right)}}\left( {\omega ,\beta ,y,\epsilon } \right) = ( &{F^{(3)}_{0,1,0,0,1,0,1}},{F^{(3)}_{ - 1,0,0,0,1,1,0}},{F^{(3)}_{0,0,0,0,1,1,0}},{F^{(3)}_{1,0,0,0,1,1,0}},{F^{(3)}_{0,1,0,0,1,1,1}},\nonumber\\
    &{F^{(3)}_{ - 1,1,0,0,1,1,1}},{F^{(3)}_{0,1,0, - 1,1,1,1}},{F^{(3)}_{ - 1,1,0,1,1,0,1}},{F^{(3)}_{0,1,0,1,1,0,1}} )^T \,.
\end{align}

To obtain the canonical basis, we have conducted a linear transformation on  $\vec f^{(i)}$, i.e., $\vec g^{(i)}( {\epsilon ,\beta ,y}) =\hat T^{(i)}( {\epsilon ,\beta ,y,\omega } )\vec f^{(i)}( {\epsilon ,\beta ,y,\omega })$,
using the  {\tt Libra} package \cite{Lee:2020zfb}.
The differential equation of $\vec g^{(i)}$ is given by
\begin{align}
   \frac{ {\partial }{{\vec g}^{\left( i \right)}}\left( {\beta ,y,\epsilon} \right) }{\partial \beta}
    = \epsilon \left( {\frac{{{{\hat a}^{\left( i \right)}}}}{{\beta  - 1}} + \frac{{{{\hat b}^{\left( i \right)}}}}{\beta } + \frac{{{{\hat c}^{\left( i \right)}}}}{{\beta  + 1}} + \frac{{{{\hat d}^{\left( i \right)}}}}{{\beta  - 1/y}} + \frac{{{{\hat e}^{\left( i \right)}}}}{{\beta  + 1/y}}} \right){{\vec g}^{\left( i \right)}}\left( {\beta ,y,\epsilon} \right)\,.
    \label{family differential equation}
\end{align} 
The explicit forms of the canonical basis and  constant matrices in the numerators are collected in a supplementary file.

As in the case at NLO, we choose the point $\beta = 0$  as the boundary. 
The regularity condition 
requires that the bases have the following structures 
\begin{align}
    \vec g^{(1)}\left( {0,y,\epsilon} \right) =  & \left({g_1^{(1)}},0,0,0,0,0, \right.\nonumber\\
    &\left. - \frac{{12}}{y}{g_1^{(1)}},6{g_1^{(1)}},{g_9^{(1)}},0, \frac{1}{y}{g_1^{(1)}} - \frac{1}{4y}{g_9^{(1)}},0,\frac{4}{{{y^2}}}{g_1^{(1)}} + \frac{2}{y^2}{g_9^{(1)}} \right)^T \,, \\
    \vec g^{(2)}\left( {0,y,\epsilon} \right)  =& \left( {{g_1^{(2)}},0,2{g_1^{(2)}}, - 2{g_1^{(2)}},0, - 2{g_1^{(2)}},0,0,0,0, - 24{g_1^{(2)}},{g_1^{(2)}}}\right)^T \,, \\
    \vec g^{(3)}\left( {0,y,\epsilon} \right) = &\left({g^{(3)}_1},{g^{(3)}_2},0, - \frac{{96}}{{3 + 2y}}{g^{(3)}_2},\frac{320}{y^2(3-2y)}\left(\frac{{2}}{3}{g^{(3)}_1} - {g^{(3)}_2}\right),\right.\nonumber \\
    &\left.0,\frac{{160}}{{y\left( {3 + 2y} \right)}}\left(- \frac{{2}}{3}{g^{(3)}_1} + {g^{(3)}_2}\right),\frac{16({1 + y - {y^2}})}{{3y^2(3+2y)}}{g^{(3)}_1},0 \right)^T \,.
\end{align}
After  performing the phase space and loop integrations, we obtain the results of the boundary integrals,
\begin{align}
    g_1^{\left( 1 \right)}\left( {0,y,\epsilon} \right) =& \frac{{4\Gamma \left( {4 - 4\epsilon } \right)}}{{{\omega ^{3 - 4\epsilon }}{\pi ^{2 - 2\epsilon }}{\Gamma ^2}\left( {1 - \epsilon } \right)}}\int {\left[ {d{k_1}} \right]\left[ {d{k_2}} \right]} \delta \left( {\omega  - {v_0} \cdot {k_1} - {v_0} \cdot {k_2}} \right) = 1 \, , \\
    g_9^{\left( 1 \right)}\left( {0,y,\epsilon} \right) =& \frac{{4\Gamma \left( {3 - 4\epsilon } \right)}}{{{\omega ^{1 - 4\epsilon }}{\pi ^{2 - 2\epsilon }}{\Gamma ^2}\left( { - \epsilon } \right)\left( {1 - 2\epsilon } \right)}}\int {\left[ {d{k_1}} \right]} \left[ {d{k_2}} \right]\frac{{\delta \left( {\omega  - {v_0} \cdot {k_1} - {v_0} \cdot {k_2}} \right)}}{{\left[ {{v_1} \cdot \left( {{k_1} + {k_2}} \right)} \right]\left( {{v_3} \cdot {k_1}} \right)}}\nonumber\label{hyp} \\
    =& \frac{{4{\epsilon ^2}}}{{\left( {1 - 2\epsilon } \right)\left( {1 - 3\epsilon } \right)}} {{\mathrm{_3F_2}}\left( {1,1 - 2\epsilon ,1 - \epsilon ;2 - 3\epsilon ,2 - 2\epsilon ;1} \right)} \,, \\
    g_1^{\left( 2 \right)}\left( {0,y,\epsilon} \right) =& g_1^{\left( 1 \right)}\left( {0,y,\epsilon} \right) = 1 \,, \\
    {g_1^{(3)}}\left( {0,y,\epsilon} \right) =&\frac{i}{{{\pi ^{3 - 2\epsilon }}{\omega ^{1 - 4\epsilon }}{e^{2i\pi \epsilon }}}}\frac{\epsilon{\left( {1 - 4\epsilon } \right)\Gamma \left( {1 - 2\epsilon } \right)}}{{\Gamma \left( {2\epsilon } \right){\Gamma ^2}\left( {1 - \epsilon } \right)}}\int {\left[ {dk} \right]{d^d}l} \frac{{\delta \left( {\omega  - {v_0} \cdot k} \right)}}{{\left( {{v_1} \cdot l} \right)\left[ {{v_3} \cdot \left( {k - l} \right)} \right]{{\left( {k - l} \right)}^2}}}\nonumber\\
    =&-e^{-2i \pi \epsilon}\frac{\Gamma(1-3\epsilon)\Gamma^2(1-2\epsilon)\Gamma(1+\epsilon)}{\Gamma(1-4\epsilon)\Gamma^2(1-\epsilon)}\,, \\
    g_2^{\left( 3 \right)}\left( {0,y,\epsilon} \right) =& - \frac{i}{{{\pi ^{3 - 2\epsilon }}{\omega ^{2 - 4\epsilon }}{e^{2i\pi \epsilon }}}}\frac{{\left( {1 - 2\epsilon } \right)\Gamma \left( {2 - 2\epsilon } \right)}}{{\Gamma \left( {2\epsilon } \right){\Gamma ^2}\left( {1 - \epsilon } \right)}}\int {\left[ {dk} \right]{d^d}l} \frac{{\delta \left( {\omega  - {v_0} \cdot k} \right)}}{{\left[ {{v_3} \cdot \left( {k - l} \right)} \right]{l^2}}} = 1 \,,
\end{align}
where the hypergeometric function in eq.~(\ref{hyp}) can be expanded order-by-order in $\epsilon$ by using the package {\tt HypExp}~\cite{Huber:2007dx}. 
These results agree with those in ref.~\cite{Wang:2018vgu}~\footnote{There is a typo in ref. \cite{Wang:2018vgu} for the result of the integral in eq.~(\ref{hyp}).}. 

With these boundary values, it is ready to solve the differential equation (\ref{family differential equation}) and the canonical bases $\vec g^{(i)}$ can be expressed in terms of MPLs.
Note that there are still two master integrals which do not depend on $\beta$. 
Thus, they cannot be calculated using differential equations. 
Instead, they can be evaluated directly,
\begin{align}
    {F^{(1)}_{1,0,1,0,0,1}} &= \int {\left[ {d{k_1}} \right]\left[ {d{k_2}} \right]} \frac{\delta \left( {\omega  - {v_0} \cdot {k_1} - {v_0} \cdot {k_2}} \right)}{{\left( {{v_1} \cdot {k_2}} \right)\left( {{v_2} \cdot {k_1}} \right)}}\frac{1}{{{{\left( {{k_1} + {k_2}} \right)}^2}}}\nonumber\\
     &  = {\pi ^{2 - 2\epsilon }}{\omega ^{ - 1 - 4\epsilon }}\frac{1}{{{{\left( {2\epsilon } \right)}^2}}}\frac{{{\Gamma ^2}\left( { - 2\epsilon } \right)\Gamma \left( { - \epsilon } \right)}}{{\Gamma \left( { - 3\epsilon } \right)\Gamma \left( { - 4\epsilon } \right)}} {\mathrm{_3F_2}\left( { - \epsilon , - \epsilon , - \epsilon ;1 - \epsilon , - 3\epsilon ;1} \right)} \,,\\
    {F^{(3)}_{0,1,1,0,0,1,1}} &=\int {\left[ {dk} \right]{d^d}l} \frac{{\delta \left( {\omega  - {v_0} \cdot k} \right)}}{{\left( {{v_1} \cdot l} \right)\left[ {{v_2} \cdot \left( {k - l} \right)} \right]}}\frac{1}{{{l^2}{{\left( {k - l} \right)}^2}}}\nonumber\\
    &=  - \frac{{{ i \pi ^{3 - 2\epsilon }}{\omega ^{ - 1 - 4\epsilon }}{e^{i\pi \epsilon }}}}{2}\frac{{\Gamma \left( {1 + \epsilon } \right)\Gamma \left( \epsilon  \right){\Gamma ^2}\left( { - \epsilon } \right)\Gamma \left( { - 2\epsilon } \right)}}{{\Gamma \left( { - 4\epsilon } \right)}}\,.
\end{align}
For the first integral, we have utilized the trick to insert a $\delta^{(d)}(q-k_1-k_2)$ function as suggested in ref. \cite{Becher:2012za}.

The NNLO soft function in Laplace space is given by
\begin{align}
\tilde s_{\rm bare}^{(2)}\left( {t,\mu ,\beta ,y} \right) = \Gamma \left( { - 4\epsilon } \right)\left( \frac{\mu^2} {s_{12}t^2}\right)^{2\epsilon} s_{\rm bare}^{(2)}\left(\beta,y  \right)\,.
\end{align}
All the divergences can be removed after performing renormalization according to eq.~(\ref{NNLO ren}).  
We have also checked that the renormalized soft function ${{\tilde s}}\left( {L,\beta,y} \right)$ satisfies the scale evolution equation (\ref{eq:gammas}).
The analytical expression of ${{\tilde s}^{\left( 2 \right)}}\left( {L,\beta,y} \right)$ is quite lengthy and can be found in the auxiliary file.

\section{Asymptotic expansions and numerical results}\label{sec:expansion}

In the limit of $\beta \to 0$,
the NLO renormalized soft function reads
\begin{align}
    {{\tilde s}^{\left( 1 \right)}}\left( {L,\beta \to 0 ,y} \right) =L^2(\mathrm{C_A}+\mathrm{C_F}) -2L\mathrm{C_F}  +\frac{\pi ^2}{6} (\mathrm{C_A}+\mathrm{C_F})+4 \mathrm{C_F}+
    \mathcal{O}(\beta)\,.
\end{align}
And the NNLO soft function is given by
\begin{align}
    &{{\tilde s}^{\left( 2 \right)}}\left( {L,\beta  \to 0,y} \right) = L^4\frac{(\mathrm{C_A} + \mathrm{C_F})^2}{2} + 
    L^3 \Bigg[ -\frac{11}{9} \mathrm{C_A^2}- 2 \mathrm{C_F^2} - \frac{29}{9} \mathrm{C_A} \mathrm{C_F}  + \frac{4}{9} (\mathrm{C_A} + \mathrm{C_F}) n_f \mathrm{T_F} \Bigg] \nonumber\\
    &+ L^2 \Bigg[\mathrm{C_A^2} \left( \frac{67}{9} - \frac{\pi^2}{6} \right) + 
    \mathrm{C_F^2} \left( 6 + \frac{\pi^2}{6} \right) +\frac{136}{9} \mathrm{C_A} \mathrm{C_F}   - \frac{4 }{9} (5\mathrm{C_A}  +8 \mathrm{C_F} )n_f \mathrm{T_F} \Bigg] \nonumber\\
    &+ L \Bigg[ \mathrm{C_A^2} \left( -\frac{404}{27} + 14 \zeta_3 \right)+\mathrm{C_F^2} \left( -8 - \frac{\pi^2}{3} \right)  + 
    \mathrm{C_A} \mathrm{C_F} \left( -\frac{1094}{27} + \frac{\pi^2}{3} + 10 \zeta_3 \right)\nonumber\\
    &+ \frac{8 }{27} (14\mathrm{C_A}  + 47\mathrm{C_F} )n_f \mathrm{T_F} \Bigg]+ \mathrm{C_A^2} \left( \frac{1214}{81} + \frac{67 \pi^2}{108} - \frac{11 \pi^4}{72} - \frac{11 \zeta_3}{9} \right)  \nonumber\\
    &+\mathrm{C_F^2} \left( 8 + \frac{2 \pi^2}{3} + \frac{\pi^4}{72} \right)+  
    \mathrm{C_A} \mathrm{C_F} \left( \frac{5918}{81} + \frac{211 \pi^2}{108} - \frac{\pi^4}{15} - \frac{101 \zeta_3}{9} \right)\nonumber\\
    &- \frac{1}{81} \Bigg[ \mathrm{C_A} \left( 328 + 15 \pi^2 - 36 \zeta_3 \right) + 
    \mathrm{C_F} \left( 2248 + 15 \pi^2 - 36 \zeta_3 \right) \Bigg]n_f \mathrm{T_F}+\mathcal{O}(\beta)\,.
\end{align}
No singular $1/\beta$ or $\log\beta$ terms appear, which agrees with the expectation that  
no new divergences, e.g., infrared or Coulomb divergences, arise when taking $\beta\to 0$.
In addition, we find that there is no dependence on $y$ in this limit because the top quark and $W$ boson are produced at rest in the center-of-mass frame of the incoming partons.

On the other hand, in the limit of $\beta \to 1$,
the top quark is highly boosted.
Consequently, new collinear divergences would appear
and are embodied in the form of $\ln^i(1-\beta)$.
The NLO soft function becomes
\begin{align}
    &{{\tilde s}^{\left( 1 \right)}}\left( {L,\beta  \to 1,y} \right)=  - \mathrm{C_F} \ln^2(1 - \beta) - 2\mathrm{C_F}\left( L+1 -\ln2 \right) \ln(1 - \beta)+L^2(\mathrm{C_A+C_F})   \nonumber\\
    &+2L\Big[-\mathrm{C_F}(1+\ln2)+(\mathrm{2C_F-C_A})\ln(1-y)+\mathrm{C_A}\ln(1+y)\Big]+ (\mathrm{C_A} - \mathrm{C_F}) \frac{5 \pi^2}{6}  \nonumber\\
    &+ \mathrm{C_F} \left( 2\ln2 - \ln^22 \right)+ \left( 2 \mathrm{C_F} - \mathrm{C_A} \right) \ln^2(1 - y)+ \mathrm{C_A} \ln^2(1 + y) - 2 \mathrm{C_A} \ln(1 - y) \ln(1 + y)\nonumber\\
    &+ 4(\mathrm{C_A} - \mathrm{C_F}) \left[ - \mathrm{Li_2}\left( \frac{1}{1 - y} \right) - 
     \mathrm{Li_2}(1 + y) + \mathrm{Li_2}\left( \frac{1 + y}{1 - y} \right) \right]
     +\mathcal{O}(1-\beta)\,,
     \label{eq:softbeta1}
\end{align}
where we have converted MPLs into the dilogarithm   and logarithm functions.
We also note the singularities at $y=\pm 1$ in the above result.
The $y=-1$ singularity corresponds to the phase space where the final-state top quark becomes collinear with the initial-state gluon. 
In this case, 
one should take it as a contribution from a collinear parton splitting convoluted with the partonic process $\bar{t}/t+b/\bar{b}\to W^-/W^+$,
which means that the top quark must be considered as a parton inside the colliding proton.
The $y=1$ singularity actually exists only at a superficial level.
It appears from the expansion of the terms like $1/\epsilon^2 \times (1-y)^{-\epsilon}$.
No such singularity remains if we take the limit $y\to 1$ firstly\footnote{This argument cannot be extended to the $y=-1$ singularity because the  hard function contains the $1/(1+y)$ pole.}.
We emphasize that these singularities appear only at the corner of $\beta=1$ {\it and} $y=\pm 1$.
Thus they can be neglected when we focus on the phenomenology at a  real collider.

The NNLO soft function in the $\beta\to 1$ limit can be found in the auxiliary file. 
Below we only present the terms with $\ln^i(1-\beta),i=4,3,2,$ to exhibit its structure:
\begin{align}
    &{{\tilde s}^{\left( 2 \right)}}\left( {L,\beta  \to 1,y} \right)=\frac{\mathrm{C_F^2}}{2} \ln^4(1-\beta)+ \mathrm{C_F}\Big[ \frac{11}{9} \mathrm{C_A} + 2 \mathrm{C_F}(L+1-\ln2) - \frac{4}{9}  n_f \mathrm{T_F}\Big] \ln^3(1-\beta)  \nonumber\\
    &+\mathrm{C_F}\Bigg\{L^2(\mathrm{C_F-C_A})+L\Big[\frac{11}{3}\mathrm{C_A}+2\mathrm{C_F}(3-\ln2)-\frac{4}{3}n_f\mathrm{T_F}+2(\mathrm{C_A-2C_F})\ln(1-y)\nonumber\\
    &-2\mathrm{C_A}\ln(1+y)\Big]+\mathrm{C_A}\bigg[-\frac{34}{9}-\frac{\pi^2}{2}-\frac{11}{3}\ln2-\ln^2(1+y)+2\ln(1-y)\ln(1+y)\bigg]\nonumber\\
    &+\mathrm{C_F}\Big(2+\frac{5\pi^2}{6}-6\ln2+3\ln^2 2\Big)+\frac{4}{9}n_f\mathrm{T_F}\Big(2+3\ln2\Big)+(\mathrm{C_A-2C_F})\ln^2(1-y)\nonumber\\
    &+4(\mathrm{C_A-C_F})\bigg[\mathrm{Li_2}\left( \frac{1}{1 - y} \right)+ 
    \mathrm{Li_2}(1 + y) - \mathrm{Li_2}\left( \frac{1 + y}{1 - y} \right)\bigg]\Bigg\}\ln^2(1-\beta)+\cdots \,.
    \label{eq:softnnlobeta1}
\end{align}

\begin{figure}[h]
    \centering
    \includegraphics[width=0.49\linewidth]{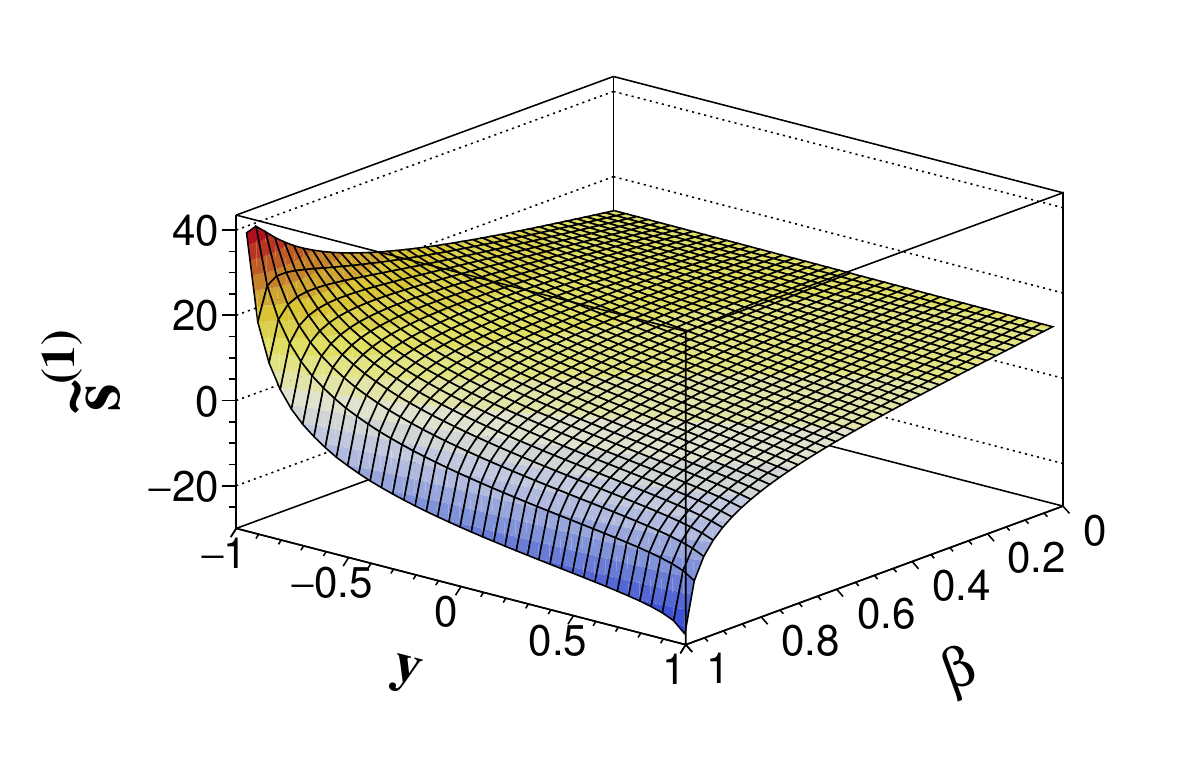}
    \includegraphics[width=0.49\linewidth]{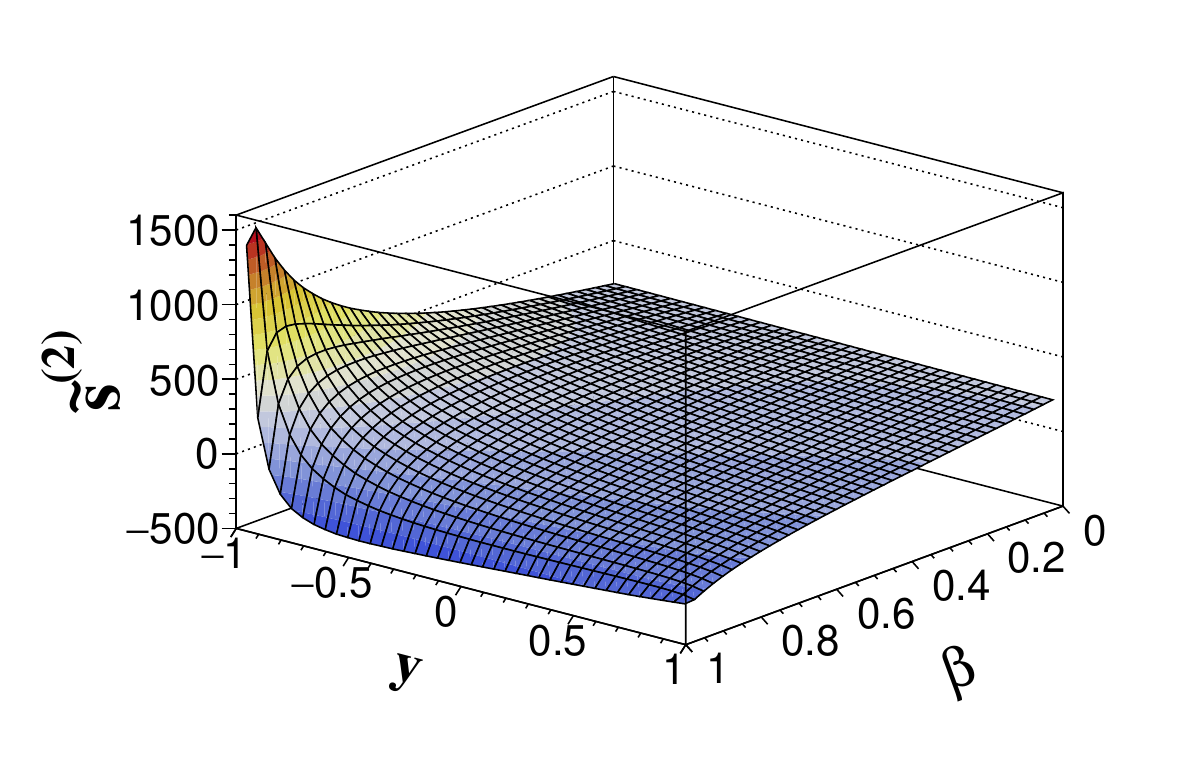}
    \caption{Pure NLO (left) and pure NNLO (right) soft functions as a function of $\beta$ and $y$.}
    \label{fig:3}
\end{figure}

In figure \ref{fig:3}, we show the soft function as a function of the velocity $\beta$ and the angular parameter  $y$.
The numerical result is calculated using $n_f=5, \mathrm{T_F}=1/2$ and $L=0$.
It can be seen that the soft function varies slowly in the region of $\beta<0.8$ and $y>-0.5$.
When $y$ becomes smaller, the soft function increases.
When $\beta $ approaches $1$, the soft function changes dramatically due to the structure shown in eqs.~(\ref{eq:softbeta1}) and (\ref{eq:softnnlobeta1}).
The NLO and NNLO soft functions exhibit similar shapes.

\begin{figure}[h]
    \centering
    \includegraphics[width=0.46\linewidth]{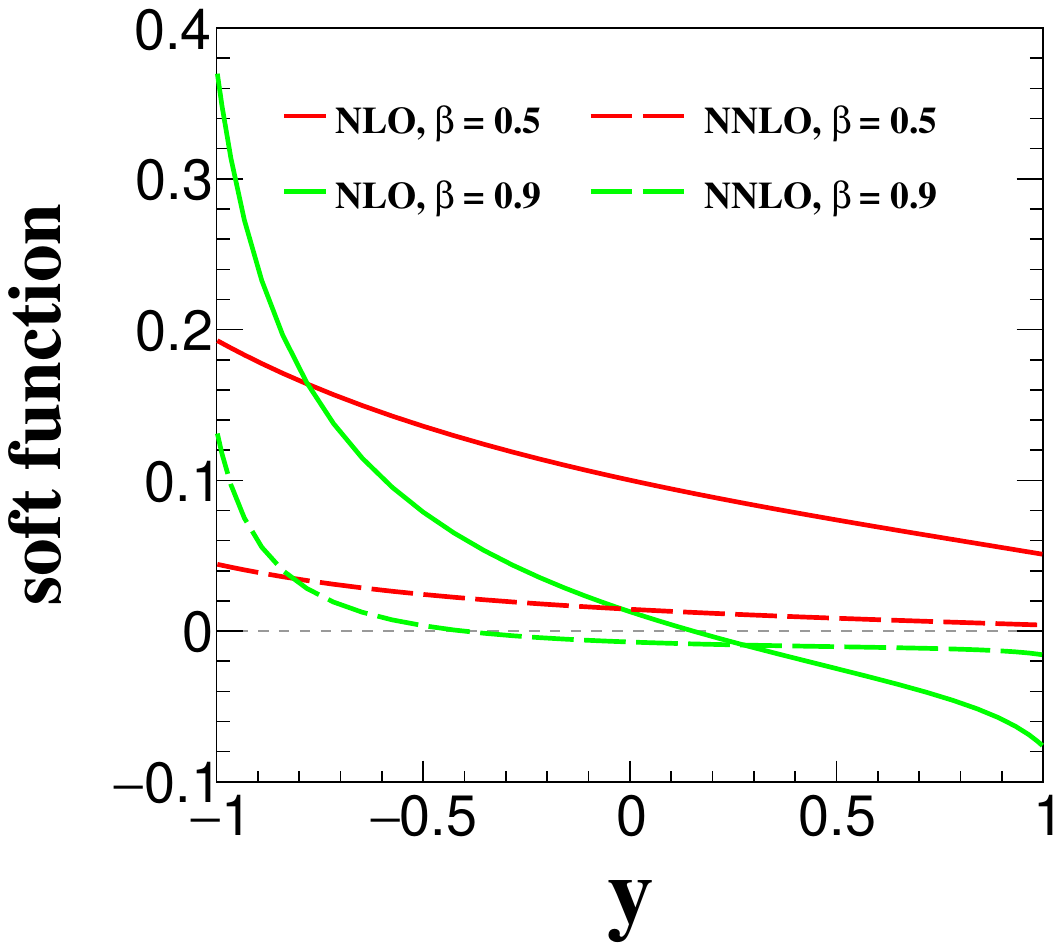}
    \includegraphics[width=0.46\linewidth]{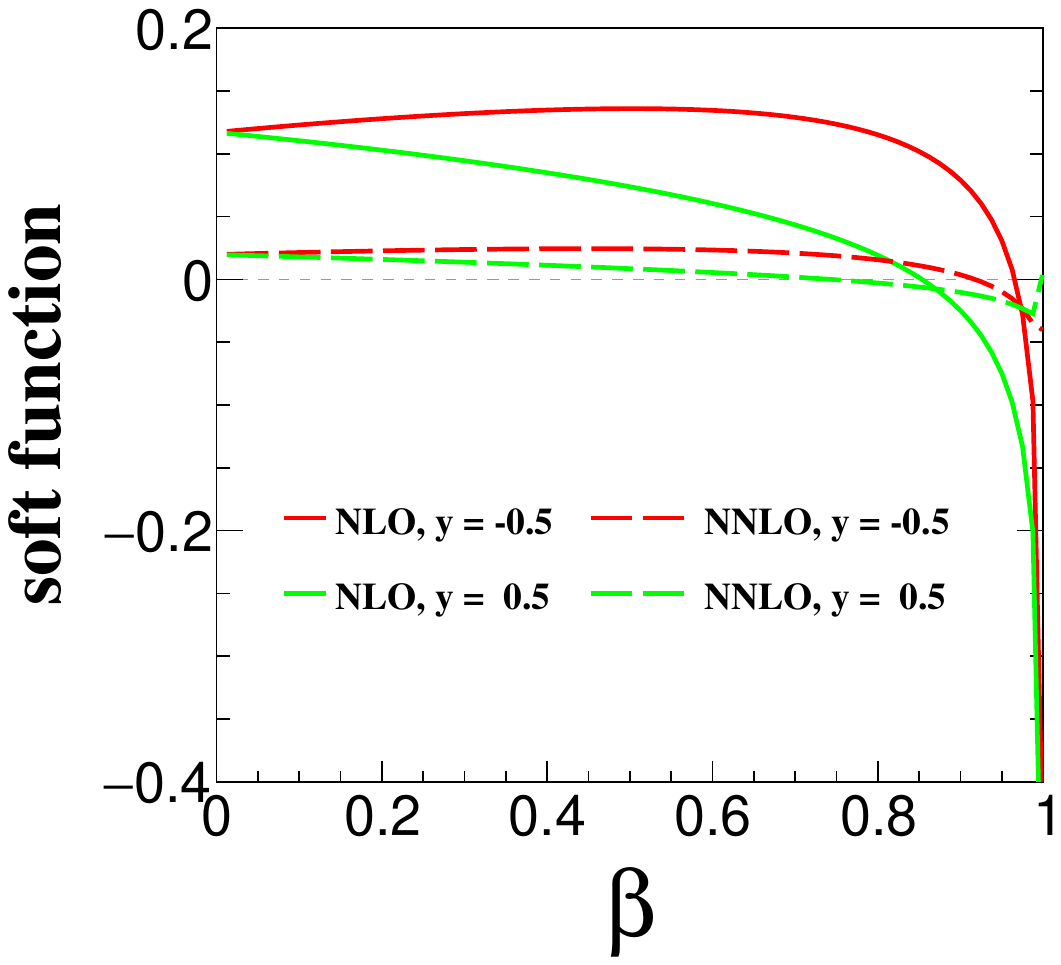}
    \caption{Pure NLO  and pure NNLO  soft functions, corresponding to eq.~(\ref{NLO ren}) and  eq.~(\ref{NNLO ren}), respectively, at fixed values of $\beta$ (left) or $y$ (right). }
    \label{fig:4}
\end{figure}

To see the impact of the soft function more clearly, we show the 2-dimensional plots with fixed values of $\beta$ or $y$ in figure  \ref{fig:4}.
We have multiplied the expansion parameter $(\alpha_s/4\pi)^i,i=1,2$, for the NLO and NNLO soft functions, respectively,
and taken $\alpha_s=0.118$ in numerical calculation. 
In the case of $\beta=0.5$, the NLO soft function changes from 0.2 to 0.05 as $y$ varies from $-1$ to $1$,
and the NNLO soft function provides a correction of about $23\%-8\%$ depending on the value of $y$. 
When setting $\beta=0.9$, we see a significant change in the NLO soft function; it changes from positive to negative values with the increasing of $y$.
The NNLO soft function has a similar distribution but the magnitude is much smaller.

We have also fixed the value of $y$ to demonstrate the variation with respect to $\beta$.
For $y=-0.5$, the soft function is stable against the change of $\beta$ except for the region near $\beta=1$, and the NNLO correction is about $17\%$ relative to the NLO result. 
For $y=0.5$, the soft function decreases fast and becomes negative around $\beta=0.85$.

\section{Conclusions}\label{sec:conclu}

In this paper, we compute the two-loop soft function for associated top quark and $W$ boson  production near threshold, which is a critical ingredient for achieving next-to-next-to-next-to-leading logarithmic  resummation of soft-gluon effects. 
We have employed IBP reduction and the method of differential equations to calculate the master integrals.
The divergences completely cancel out after performing the renormalization using the known anomalous dimension of the soft function, which validates our calculation.
The full analytic results of the NLO and NNLO 
soft functions are expressed in terms of MPLs.
The asymptotic behaviors of the soft function in the low-energy limit ($\beta \to 0$) and the high-energy limit ($\beta \to 1$) are also analyzed.
Numerical results demonstrate the stability of the soft function in moderate kinematic regions
and the rapid change in the high energy limit.
Our results show that the NNLO corrections are around 
 $10\%-20\%$ with respect to the NLO soft function in the major part of the phase space.

\section*{Acknowledgement}
This work was partly supported by the National Natural Science Foundation of China under grant No. 12275156, No. 12321005 and the Taishan Scholar Foundation of Shandong province (tsqn201909011).
The Feynman diagrams were drawn using {\tt JaxoDraw}~\cite{Binosi:2008ig}.

\bibliographystyle{JHEP}
\bibliography{ref}

\providecommand{\href}[2]{#2}\begingroup\raggedright\begin{thebibliography}{10}

\bibitem{ATLAS:2012bqt}
{\scshape ATLAS} collaboration, G.~Aad et~al., \emph{{Evidence for the
  associated production of a $W$ boson and a top quark in ATLAS at $\sqrt{s}=7$
  TeV}}, \href{https://doi.org/10.1016/j.physletb.2012.08.011}{\emph{Phys.
  Lett. B} {\bfseries 716} (2012) 142--159},
  [\href{https://arxiv.org/abs/1205.5764}{{\ttfamily 1205.5764}}].

\bibitem{CMS:2012pxd}
{\scshape CMS} collaboration, S.~Chatrchyan et~al., \emph{{Evidence for
  Associated Production of a Single Top Quark and W Boson in $pp$ Collisions at
  $\sqrt{s}$ = 7 TeV}},
  \href{https://doi.org/10.1103/PhysRevLett.110.022003}{\emph{Phys. Rev. Lett.}
  {\bfseries 110} (2013) 022003},
  [\href{https://arxiv.org/abs/1209.3489}{{\ttfamily 1209.3489}}].

\bibitem{CMS:2014fut}
{\scshape CMS} collaboration, S.~Chatrchyan et~al., \emph{{Observation of the
  associated production of a single top quark and a $W$ boson in $pp$
  collisions at $\sqrt s = $8 TeV}},
  \href{https://doi.org/10.1103/PhysRevLett.112.231802}{\emph{Phys. Rev. Lett.}
  {\bfseries 112} (2014) 231802},
  [\href{https://arxiv.org/abs/1401.2942}{{\ttfamily 1401.2942}}].

\bibitem{ATLAS:2015igu}
{\scshape ATLAS} collaboration, G.~Aad et~al., \emph{{Measurement of the
  production cross-section of a single top quark in association with a $W$
  boson at 8 TeV with the ATLAS experiment}},
  \href{https://doi.org/10.1007/JHEP01(2016)064}{\emph{JHEP} {\bfseries 01}
  (2016) 064}, [\href{https://arxiv.org/abs/1510.03752}{{\ttfamily
  1510.03752}}].

\bibitem{ATLAS:2016ofl}
{\scshape ATLAS} collaboration, M.~Aaboud et~al., \emph{{Measurement of the
  cross-section for producing a W boson in association with a single top quark
  in pp collisions at $ \sqrt{s}=13 $ TeV with ATLAS}},
  \href{https://doi.org/10.1007/JHEP01(2018)063}{\emph{JHEP} {\bfseries 01}
  (2018) 063}, [\href{https://arxiv.org/abs/1612.07231}{{\ttfamily
  1612.07231}}].

\bibitem{ATLAS:2017quy}
{\scshape ATLAS} collaboration, M.~Aaboud et~al., \emph{{Measurement of
  differential cross-sections of a single top quark produced in association
  with a $W$ boson at $\sqrt{s}=13$ TeV with ATLAS}},
  \href{https://doi.org/10.1140/epjc/s10052-018-5649-8}{\emph{Eur. Phys. J. C}
  {\bfseries 78} (2018) 186},
  [\href{https://arxiv.org/abs/1712.01602}{{\ttfamily 1712.01602}}].

\bibitem{CMS:2018amb}
{\scshape CMS} collaboration, A.~M. Sirunyan et~al., \emph{{Measurement of the
  production cross section for single top quarks in association with W bosons
  in proton-proton collisions at $ \sqrt{s}=13 $ TeV}},
  \href{https://doi.org/10.1007/JHEP10(2018)117}{\emph{JHEP} {\bfseries 10}
  (2018) 117}, [\href{https://arxiv.org/abs/1805.07399}{{\ttfamily
  1805.07399}}].

\bibitem{CMS:2022ytw}
{\scshape CMS} collaboration, A.~Tumasyan et~al., \emph{{Measurement of
  inclusive and differential cross sections for single top quark production in
  association with a W boson in proton-proton collisions at $ \sqrt{s} $ = 13
  TeV}}, \href{https://doi.org/10.1007/JHEP07(2023)046}{\emph{JHEP} {\bfseries
  07} (2023) 046}, [\href{https://arxiv.org/abs/2208.00924}{{\ttfamily
  2208.00924}}].

\bibitem{ATLAS:2024ppp}
{\scshape ATLAS} collaboration, G.~Aad et~al., \emph{{Measurement of single
  top-quark production in association with a W boson in pp collisions at
  s=13\,\,TeV with the ATLAS detector}},
  \href{https://doi.org/10.1103/PhysRevD.110.072010}{\emph{Phys. Rev. D}
  {\bfseries 110} (2024) 072010},
  [\href{https://arxiv.org/abs/2407.15594}{{\ttfamily 2407.15594}}].

\bibitem{CMS:2024okz}
{\scshape CMS} collaboration, A.~Hayrapetyan et~al., \emph{{Measurement of
  inclusive and differential cross sections of single top quark production in
  association with a W boson in proton-proton collisions at $ \sqrt{s} $ = 13.6
  TeV}}, \href{https://doi.org/10.1007/JHEP01(2025)107}{\emph{JHEP} {\bfseries
  01} (2025) 107}, [\href{https://arxiv.org/abs/2409.06444}{{\ttfamily
  2409.06444}}].

\bibitem{ATLAS:2018ivx}
{\scshape ATLAS} collaboration, M.~Aaboud et~al., \emph{{Probing the quantum
  interference between singly and doubly resonant top-quark production in $pp$
  collisions at $\sqrt{s}=13$ TeV with the ATLAS detector}},
  \href{https://doi.org/10.1103/PhysRevLett.121.152002}{\emph{Phys. Rev. Lett.}
  {\bfseries 121} (2018) 152002},
  [\href{https://arxiv.org/abs/1806.04667}{{\ttfamily 1806.04667}}].

\bibitem{Belyaev:1998dn}
A.~S. Belyaev, E.~E. Boos and L.~V. Dudko, \emph{{Single top quark at future
  hadron colliders: Complete signal and background study}},
  \href{https://doi.org/10.1103/PhysRevD.59.075001}{\emph{Phys. Rev. D}
  {\bfseries 59} (1999) 075001},
  [\href{https://arxiv.org/abs/hep-ph/9806332}{{\ttfamily hep-ph/9806332}}].

\bibitem{Tait:1999cf}
T.~M.~P. Tait, \emph{{The $t W^{-}$ mode of single top production}},
  \href{https://doi.org/10.1103/PhysRevD.61.034001}{\emph{Phys. Rev. D}
  {\bfseries 61} (1999) 034001},
  [\href{https://arxiv.org/abs/hep-ph/9909352}{{\ttfamily hep-ph/9909352}}].

\bibitem{Campbell:2005bb}
J.~M. Campbell and F.~Tramontano, \emph{{Next-to-leading order corrections to
  Wt production and decay}},
  \href{https://doi.org/10.1016/j.nuclphysb.2005.08.015}{\emph{Nucl. Phys. B}
  {\bfseries 726} (2005) 109--130},
  [\href{https://arxiv.org/abs/hep-ph/0506289}{{\ttfamily hep-ph/0506289}}].

\bibitem{Frixione:2008yi}
S.~Frixione, E.~Laenen, P.~Motylinski, B.~R. Webber and C.~D. White,
  \emph{{Single-top hadroproduction in association with a W boson}},
  \href{https://doi.org/10.1088/1126-6708/2008/07/029}{\emph{JHEP} {\bfseries
  07} (2008) 029}, [\href{https://arxiv.org/abs/0805.3067}{{\ttfamily
  0805.3067}}].

\bibitem{Demartin:2016axk}
F.~Demartin, B.~Maier, F.~Maltoni, K.~Mawatari and M.~Zaro, \emph{{tWH
  associated production at the LHC}},
  \href{https://doi.org/10.1140/epjc/s10052-017-4601-7}{\emph{Eur. Phys. J. C}
  {\bfseries 77} (2017) 34},
  [\href{https://arxiv.org/abs/1607.05862}{{\ttfamily 1607.05862}}].

\bibitem{Dong:2024jxp}
L.~Dong, H.~T. Li, Z.-Y. Li and J.~Wang, \emph{{Subtraction of the $
  t\overline{t} $ contribution in $ tW\overline{b} $ production at the one-loop
  level}}, \href{https://doi.org/10.1007/JHEP01(2025)158}{\emph{JHEP}
  {\bfseries 01} (2025) 158},
  [\href{https://arxiv.org/abs/2411.07455}{{\ttfamily 2411.07455}}].

\bibitem{Giele:1995kr}
W.~T. Giele, S.~Keller and E.~Laenen, \emph{{QCD corrections to $W$ boson plus
  heavy quark production at the Tevatron}},
  \href{https://doi.org/10.1016/0370-2693(96)00078-0}{\emph{Phys. Lett. B}
  {\bfseries 372} (1996) 141--149},
  [\href{https://arxiv.org/abs/hep-ph/9511449}{{\ttfamily hep-ph/9511449}}].

\bibitem{Cao:2008af}
Q.-H. Cao, \emph{{Demonstration of One Cutoff Phase Space Slicing Method:
  Next-to-Leading Order QCD Corrections to the tW Associated Production in
  Hadron Collision}},  \href{https://arxiv.org/abs/0801.1539}{{\ttfamily
  0801.1539}}.

\bibitem{Kant:2014oha}
P.~Kant, O.~M. Kind, T.~Kintscher, T.~Lohse, T.~Martini, S.~M\"olbitz et~al.,
  \emph{{HatHor for single top-quark production: Updated predictions and
  uncertainty estimates for single top-quark production in hadronic
  collisions}}, \href{https://doi.org/10.1016/j.cpc.2015.02.001}{\emph{Comput.
  Phys. Commun.} {\bfseries 191} (2015) 74--89},
  [\href{https://arxiv.org/abs/1406.4403}{{\ttfamily 1406.4403}}].

\bibitem{Kidonakis:2006bu}
N.~Kidonakis, \emph{{Single top production at the Tevatron: Threshold
  resummation and finite-order soft gluon corrections}},
  \href{https://doi.org/10.1103/PhysRevD.74.114012}{\emph{Phys. Rev. D}
  {\bfseries 74} (2006) 114012},
  [\href{https://arxiv.org/abs/hep-ph/0609287}{{\ttfamily hep-ph/0609287}}].

\bibitem{Kidonakis:2010ux}
N.~Kidonakis, \emph{{Two-loop soft anomalous dimensions for single top quark
  associated production with a $W^-$ or $H^-$}},
  \href{https://doi.org/10.1103/PhysRevD.82.054018}{\emph{Phys. Rev. D}
  {\bfseries 82} (2010) 054018},
  [\href{https://arxiv.org/abs/1005.4451}{{\ttfamily 1005.4451}}].

\bibitem{Kidonakis:2016sjf}
N.~Kidonakis, \emph{{Soft-gluon corrections for $tW$ production at N$^3$LO}},
  \href{https://doi.org/10.1103/PhysRevD.96.034014}{\emph{Phys. Rev. D}
  {\bfseries 96} (2017) 034014},
  [\href{https://arxiv.org/abs/1612.06426}{{\ttfamily 1612.06426}}].

\bibitem{Kidonakis:2021vob}
N.~Kidonakis and N.~Yamanaka, \emph{{Higher-order corrections for $tW$
  production at high-energy hadron colliders}},
  \href{https://doi.org/10.1007/JHEP05(2021)278}{\emph{JHEP} {\bfseries 05}
  (2021) 278}, [\href{https://arxiv.org/abs/2102.11300}{{\ttfamily
  2102.11300}}].

\bibitem{Li:2019dhg}
C.~S. Li, H.~T. Li, D.~Y. Shao and J.~Wang, \emph{{Momentum-space threshold
  resummation in $tW$ production at the LHC}},
  \href{https://doi.org/10.1007/JHEP06(2019)125}{\emph{JHEP} {\bfseries 06}
  (2019) 125}, [\href{https://arxiv.org/abs/1903.01646}{{\ttfamily
  1903.01646}}].

\bibitem{Re:2010bp}
E.~Re, \emph{{Single-top Wt-channel production matched with parton showers
  using the POWHEG method}},
  \href{https://doi.org/10.1140/epjc/s10052-011-1547-z}{\emph{Eur. Phys. J. C}
  {\bfseries 71} (2011) 1547},
  [\href{https://arxiv.org/abs/1009.2450}{{\ttfamily 1009.2450}}].

\bibitem{Jezo:2016ujg}
T.~Je\v{z}o, J.~M. Lindert, P.~Nason, C.~Oleari and S.~Pozzorini, \emph{{An
  NLO+PS generator for $t\bar{t}$ and $Wt$ production and decay including
  non-resonant and interference effects}},
  \href{https://doi.org/10.1140/epjc/s10052-016-4538-2}{\emph{Eur. Phys. J. C}
  {\bfseries 76} (2016) 691},
  [\href{https://arxiv.org/abs/1607.04538}{{\ttfamily 1607.04538}}].

\bibitem{Chen:2021gjv}
L.-B. Chen and J.~Wang, \emph{{Analytic two-loop master integrals for tW
  production at hadron colliders: I *}},
  \href{https://doi.org/10.1088/1674-1137/ac2a1e}{\emph{Chin. Phys. C}
  {\bfseries 45} (2021) 123106},
  [\href{https://arxiv.org/abs/2106.12093}{{\ttfamily 2106.12093}}].

\bibitem{Long:2021vse}
M.-M. Long, R.-Y. Zhang, W.-G. Ma, Y.~Jiang, L.~Han, Z.~Li et~al.,
  \emph{{Two-loop master integrals for the single top production associated
  with $W$ boson}},  \href{https://arxiv.org/abs/2111.14172}{{\ttfamily
  2111.14172}}.

\bibitem{Wang:2022enl}
J.~Wang and Y.~Wang, \emph{{Analytic two-loop master integrals for tW
  production at hadron colliders. Part II}},
  \href{https://doi.org/10.1007/JHEP02(2023)127}{\emph{JHEP} {\bfseries 02}
  (2023) 127}, [\href{https://arxiv.org/abs/2211.13713}{{\ttfamily
  2211.13713}}].

\bibitem{Chen:2022ntw}
L.-B. Chen, L.~Dong, H.~T. Li, Z.~Li, J.~Wang and Y.~Wang, \emph{{One-loop
  squared amplitudes for hadronic tW production at next-to-next-to-leading
  order in QCD}}, \href{https://doi.org/10.1007/JHEP08(2022)211}{\emph{JHEP}
  {\bfseries 08} (2022) 211},
  [\href{https://arxiv.org/abs/2204.13500}{{\ttfamily 2204.13500}}].

\bibitem{Chen:2022yni}
L.-B. Chen, L.~Dong, H.~T. Li, Z.~Li, J.~Wang and Y.~Wang, \emph{{Analytic
  two-loop QCD amplitudes for tW production: Leading color and light
  fermion-loop contributions}},
  \href{https://doi.org/10.1103/PhysRevD.106.096029}{\emph{Phys. Rev. D}
  {\bfseries 106} (2022) 096029},
  [\href{https://arxiv.org/abs/2208.08786}{{\ttfamily 2208.08786}}].

\bibitem{Chen:2022pdw}
L.-B. Chen, L.~Dong, H.~T. Li, Z.~Li, J.~Wang and Y.~Wang, \emph{{Complete
  two-loop QCD amplitudes for tW production at hadron colliders}},
  \href{https://doi.org/10.1007/JHEP07(2023)089}{\emph{JHEP} {\bfseries 07}
  (2023) 089}, [\href{https://arxiv.org/abs/2212.07190}{{\ttfamily
  2212.07190}}].

\bibitem{Li:2016tvb}
H.~T. Li and J.~Wang, \emph{{Next-to-Next-to-Leading Order $N$-Jettiness Soft
  Function for One Massive Colored Particle Production at Hadron Colliders}},
  \href{https://doi.org/10.1007/JHEP02(2017)002}{\emph{JHEP} {\bfseries 02}
  (2017) 002}, [\href{https://arxiv.org/abs/1611.02749}{{\ttfamily
  1611.02749}}].

\bibitem{Li:2018tsq}
H.~T. Li and J.~Wang, \emph{{Next-to-next-to-leading order $N$-jettiness soft
  function for $tW$ production}},
  \href{https://doi.org/10.1016/j.physletb.2018.08.019}{\emph{Phys. Lett. B}
  {\bfseries 784} (2018) 397--404},
  [\href{https://arxiv.org/abs/1804.06358}{{\ttfamily 1804.06358}}].

\bibitem{Cutkosky:1960sp}
R.~E. Cutkosky, \emph{{Singularities and discontinuities of Feynman
  amplitudes}}, \href{https://doi.org/10.1063/1.1703676}{\emph{J. Math. Phys.}
  {\bfseries 1} (1960) 429--433}.

\bibitem{Anastasiou:2002yz}
C.~Anastasiou and K.~Melnikov, \emph{{Higgs boson production at hadron
  colliders in NNLO QCD}},
  \href{https://doi.org/10.1016/S0550-3213(02)00837-4}{\emph{Nucl. Phys. B}
  {\bfseries 646} (2002) 220--256},
  [\href{https://arxiv.org/abs/hep-ph/0207004}{{\ttfamily hep-ph/0207004}}].

\bibitem{Tkachov:1981wb}
F.~V. Tkachov, \emph{{A theorem on analytical calculability of 4-loop
  renormalization group functions}},
  \href{https://doi.org/10.1016/0370-2693(81)90288-4}{\emph{Phys. Lett. B}
  {\bfseries 100} (1981) 65--68}.

\bibitem{Chetyrkin:1981qh}
K.~G. Chetyrkin and F.~V. Tkachov, \emph{{Integration by parts: The algorithm
  to calculate $\beta$-functions in 4 loops}},
  \href{https://doi.org/10.1016/0550-3213(81)90199-1}{\emph{Nucl. Phys. B}
  {\bfseries 192} (1981) 159--204}.

\bibitem{Smirnov:2019qkx}
A.~V. Smirnov and F.~S. Chukharev, \emph{{FIRE6: Feynman Integral REduction
  with modular arithmetic}},
  \href{https://doi.org/10.1016/j.cpc.2019.106877}{\emph{Comput. Phys. Commun.}
  {\bfseries 247} (2020) 106877},
  [\href{https://arxiv.org/abs/1901.07808}{{\ttfamily 1901.07808}}].

\bibitem{Kotikov:1990kg}
A.~V. Kotikov, \emph{{Differential equations method: New technique for massive
  Feynman diagrams calculation}},
  \href{https://doi.org/10.1016/0370-2693(91)90413-K}{\emph{Phys. Lett. B}
  {\bfseries 254} (1991) 158--164}.

\bibitem{Kotikov:1991pm}
A.~V. Kotikov, \emph{{Differential equation method: The Calculation of N point
  Feynman diagrams}},
  \href{https://doi.org/10.1016/0370-2693(91)90536-Y}{\emph{Phys. Lett. B}
  {\bfseries 267} (1991) 123--127}.

\bibitem{Gehrmann:1999as}
T.~Gehrmann and E.~Remiddi, \emph{{Differential equations for two-loop
  four-point functions}},
  \href{https://doi.org/10.1016/S0550-3213(00)00223-6}{\emph{Nucl. Phys. B}
  {\bfseries 580} (2000) 485--518},
  [\href{https://arxiv.org/abs/hep-ph/9912329}{{\ttfamily hep-ph/9912329}}].

\bibitem{Henn:2013pwa}
J.~M. Henn, \emph{{Multiloop integrals in dimensional regularization made
  simple}}, \href{https://doi.org/10.1103/PhysRevLett.110.251601}{\emph{Phys.
  Rev. Lett.} {\bfseries 110} (2013) 251601},
  [\href{https://arxiv.org/abs/1304.1806}{{\ttfamily 1304.1806}}].

\bibitem{2011arXiv1105.2076G}
A.~B. {Goncharov}, \emph{{Multiple polylogarithms, cyclotomy and modular
  complexes}}, \href{https://doi.org/10.48550/arXiv.1105.2076}{\emph{arXiv
  e-prints} (May, 2011) arXiv:1105.2076},
  [\href{https://arxiv.org/abs/1105.2076}{{\ttfamily 1105.2076}}].

\bibitem{Becher:2007ty}
T.~Becher, M.~Neubert and G.~Xu, \emph{{Dynamical Threshold Enhancement and
  Resummation in Drell-Yan Production}},
  \href{https://doi.org/10.1088/1126-6708/2008/07/030}{\emph{JHEP} {\bfseries
  07} (2008) 030}, [\href{https://arxiv.org/abs/0710.0680}{{\ttfamily
  0710.0680}}].

\bibitem{Beneke:2002ph}
M.~Beneke, A.~P. Chapovsky, M.~Diehl and T.~Feldmann, \emph{{Soft collinear
  effective theory and heavy to light currents beyond leading power}},
  \href{https://doi.org/10.1016/S0550-3213(02)00687-9}{\emph{Nucl. Phys. B}
  {\bfseries 643} (2002) 431--476},
  [\href{https://arxiv.org/abs/hep-ph/0206152}{{\ttfamily hep-ph/0206152}}].

\bibitem{Chay:2004zn}
J.~Chay, C.~Kim, Y.~G. Kim and J.-P. Lee, \emph{{Soft Wilson lines in
  soft-collinear effective theory}},
  \href{https://doi.org/10.1103/PhysRevD.71.056001}{\emph{Phys. Rev. D}
  {\bfseries 71} (2005) 056001},
  [\href{https://arxiv.org/abs/hep-ph/0412110}{{\ttfamily hep-ph/0412110}}].

\bibitem{Korchemsky:1991zp}
G.~P. Korchemsky and A.~V. Radyushkin, \emph{{Infrared factorization, Wilson
  lines and the heavy quark limit}},
  \href{https://doi.org/10.1016/0370-2693(92)90405-S}{\emph{Phys. Lett. B}
  {\bfseries 279} (1992) 359--366},
  [\href{https://arxiv.org/abs/hep-ph/9203222}{{\ttfamily hep-ph/9203222}}].

\bibitem{Catani:1996jh}
S.~Catani and M.~H. Seymour, \emph{{The Dipole formalism for the calculation of
  QCD jet cross-sections at next-to-leading order}},
  \href{https://doi.org/10.1016/0370-2693(96)00425-X}{\emph{Phys. Lett. B}
  {\bfseries 378} (1996) 287--301},
  [\href{https://arxiv.org/abs/hep-ph/9602277}{{\ttfamily hep-ph/9602277}}].

\bibitem{Catani:1996vz}
S.~Catani and M.~H. Seymour, \emph{{A General algorithm for calculating jet
  cross-sections in NLO QCD}},
  \href{https://doi.org/10.1016/S0550-3213(96)00589-5}{\emph{Nucl. Phys. B}
  {\bfseries 485} (1997) 291--419},
  [\href{https://arxiv.org/abs/hep-ph/9605323}{{\ttfamily hep-ph/9605323}}].

\bibitem{Ferroglia:2009ii}
A.~Ferroglia, M.~Neubert, B.~D. Pecjak and L.~L. Yang, \emph{{Two-loop
  divergences of massive scattering amplitudes in non-abelian gauge theories}},
  \href{https://doi.org/10.1088/1126-6708/2009/11/062}{\emph{JHEP} {\bfseries
  11} (2009) 062}, [\href{https://arxiv.org/abs/0908.3676}{{\ttfamily
  0908.3676}}].

\bibitem{Korchemsky:1992xv}
G.~P. Korchemsky and G.~Marchesini, \emph{{Structure function for large x and
  renormalization of Wilson loop}},
  \href{https://doi.org/10.1016/0550-3213(93)90167-N}{\emph{Nucl. Phys. B}
  {\bfseries 406} (1993) 225--258},
  [\href{https://arxiv.org/abs/hep-ph/9210281}{{\ttfamily hep-ph/9210281}}].

\bibitem{Moch:2004pa}
S.~Moch, J.~A.~M. Vermaseren and A.~Vogt, \emph{{The Three loop splitting
  functions in QCD: The Nonsinglet case}},
  \href{https://doi.org/10.1016/j.nuclphysb.2004.03.030}{\emph{Nucl. Phys. B}
  {\bfseries 688} (2004) 101--134},
  [\href{https://arxiv.org/abs/hep-ph/0403192}{{\ttfamily hep-ph/0403192}}].

\bibitem{Ahrens:2010zv}
V.~Ahrens, A.~Ferroglia, M.~Neubert, B.~D. Pecjak and L.~L. Yang,
  \emph{{Renormalization-Group Improved Predictions for Top-Quark Pair
  Production at Hadron Colliders}},
  \href{https://doi.org/10.1007/JHEP09(2010)097}{\emph{JHEP} {\bfseries 09}
  (2010) 097}, [\href{https://arxiv.org/abs/1003.5827}{{\ttfamily 1003.5827}}].

\bibitem{Laporta:2000dsw}
S.~Laporta, \emph{{High-precision calculation of multiloop Feynman integrals by
  difference equations}},
  \href{https://doi.org/10.1142/S0217751X00002159}{\emph{Int. J. Mod. Phys. A}
  {\bfseries 15} (2000) 5087--5159},
  [\href{https://arxiv.org/abs/hep-ph/0102033}{{\ttfamily hep-ph/0102033}}].

\bibitem{Duhr:2014woa}
C.~Duhr, \emph{{Mathematical aspects of scattering amplitudes}},  in
  \emph{{Theoretical Advanced Study Institute in Elementary Particle Physics}:
  {Journeys Through the Precision Frontier: Amplitudes for Colliders}},
  pp.~419--476, 2015, \href{https://arxiv.org/abs/1411.7538}{{\ttfamily
  1411.7538}}, \href{https://doi.org/10.1142/9789814678766_0010}{DOI}.

\bibitem{Bauer:2000cp}
C.~W. Bauer, A.~Frink and R.~Kreckel, \emph{{Introduction to the GiNaC
  framework for symbolic computation within the C++ programming language}},
  \href{https://doi.org/10.1006/jsco.2001.0494}{\emph{J. Symb. Comput.}
  {\bfseries 33} (2002) 1--12},
  [\href{https://arxiv.org/abs/cs/0004015}{{\ttfamily cs/0004015}}].

\bibitem{Duhr:2019tlz}
C.~Duhr and F.~Dulat, \emph{{PolyLogTools \textemdash{} polylogs for the
  masses}}, \href{https://doi.org/10.1007/JHEP08(2019)135}{\emph{JHEP}
  {\bfseries 08} (2019) 135},
  [\href{https://arxiv.org/abs/1904.07279}{{\ttfamily 1904.07279}}].

\bibitem{Catani:1999ss}
S.~Catani and M.~Grazzini, \emph{{Infrared factorization of tree level QCD
  amplitudes at the next-to-next-to-leading order and beyond}},
  \href{https://doi.org/10.1016/S0550-3213(99)00778-6}{\emph{Nucl. Phys. B}
  {\bfseries 570} (2000) 287--325},
  [\href{https://arxiv.org/abs/hep-ph/9908523}{{\ttfamily hep-ph/9908523}}].

\bibitem{Czakon:2011ve}
M.~Czakon, \emph{{Double-real radiation in hadronic top quark pair production
  as a proof of a certain concept}},
  \href{https://doi.org/10.1016/j.nuclphysb.2011.03.020}{\emph{Nucl. Phys. B}
  {\bfseries 849} (2011) 250--295},
  [\href{https://arxiv.org/abs/1101.0642}{{\ttfamily 1101.0642}}].

\bibitem{Bierenbaum:2011gg}
I.~Bierenbaum, M.~Czakon and A.~Mitov, \emph{{The singular behavior of one-loop
  massive QCD amplitudes with one external soft gluon}},
  \href{https://doi.org/10.1016/j.nuclphysb.2011.11.002}{\emph{Nucl. Phys. B}
  {\bfseries 856} (2012) 228--246},
  [\href{https://arxiv.org/abs/1107.4384}{{\ttfamily 1107.4384}}].

\bibitem{Lee:2020zfb}
R.~N. Lee, \emph{{Libra: A package for transformation of differential systems
  for multiloop integrals}},
  \href{https://doi.org/10.1016/j.cpc.2021.108058}{\emph{Comput. Phys. Commun.}
  {\bfseries 267} (2021) 108058},
  [\href{https://arxiv.org/abs/2012.00279}{{\ttfamily 2012.00279}}].

\bibitem{Huber:2007dx}
T.~Huber and D.~Maitre, \emph{{HypExp 2, Expanding Hypergeometric Functions
  about Half-Integer Parameters}},
  \href{https://doi.org/10.1016/j.cpc.2007.12.008}{\emph{Comput. Phys. Commun.}
  {\bfseries 178} (2008) 755--776},
  [\href{https://arxiv.org/abs/0708.2443}{{\ttfamily 0708.2443}}].

\bibitem{Wang:2018vgu}
G.~Wang, X.~Xu, L.~L. Yang and H.~X. Zhu, \emph{{The next-to-next-to-leading
  order soft function for top quark pair production}},
  \href{https://doi.org/10.1007/JHEP06(2018)013}{\emph{JHEP} {\bfseries 06}
  (2018) 013}, [\href{https://arxiv.org/abs/1804.05218}{{\ttfamily
  1804.05218}}].

\bibitem{Becher:2012za}
T.~Becher, G.~Bell and S.~Marti, \emph{{NNLO soft function for electroweak
  boson production at large transverse momentum}},
  \href{https://doi.org/10.1007/JHEP04(2012)034}{\emph{JHEP} {\bfseries 04}
  (2012) 034}, [\href{https://arxiv.org/abs/1201.5572}{{\ttfamily 1201.5572}}].

\bibitem{Binosi:2008ig}
D.~Binosi, J.~Collins, C.~Kaufhold and L.~Theussl, \emph{{JaxoDraw: A Graphical
  user interface for drawing Feynman diagrams. Version 2.0 release notes}},
  \href{https://doi.org/10.1016/j.cpc.2009.02.020}{\emph{Comput. Phys. Commun.}
  {\bfseries 180} (2009) 1709--1715},
  [\href{https://arxiv.org/abs/0811.4113}{{\ttfamily 0811.4113}}].

\end{thebibliography}\endgroup
\end{document}